\documentclass[a4paper,11pt]{article}
\usepackage{jinstpub} 
\usepackage{float}
\usepackage{subcaption}

\title{\boldmath The throttling refrigeration system for high cooling power recovery of the PandaX-xT cryogenic distillation system for radon removal}

\author[a,1]{Shunyu Yao,\note[1]{These authors contributed equally to this work.}}
\author[b,1]{Kangkang Zhao,}
\author[b,c,d,e,*]{Zhou Wang,\note[*]{Corresponding author.}}
\author[f]{Zhi Zheng,}
\author[c]{Haoyu Wang,}
\author[b,*]{Xiangyi Cui,}
\author[g]{Haifeng Wang,}
\author[b,c,d,e]{Tao Zhang,}
\author[b,c,d,e]{Li Zhao,}
\author[f]{Huaikuang Ding,}
\author[f]{Wenbing Tao,}
\author[h]{Xiang Xiao,}
\author[a,c,e,*]{Shaobo Wang,}
\author[i]{Yonglin Ju,}
\author[b,c,d,e]{Jianglai Liu,}
\author[j]{Xiangdong Ji,}
\author[d,k]{Shuaijie Li,}
\author[k]{Manbin Shen,}
\author[k]{Chengbo Du}

\affiliation[a]{ SJTU Paris Elite Institute of Technology, Shanghai Jiao Tong University,\\Shanghai 200240, China}
\affiliation[b]{New Cornerstone Science Laboratory, Tsung-Dao Lee Institute, Shanghai Jiao Tong University, \\Shanghai 200240, China}
\affiliation[c]{ School of Physics and Astronomy, Shanghai Jiao Tong University, Key Laboratory for Particle Astrophysics and Cosmology (MoE),Shanghai Key Laboratory for Particle Physics and Cosmology,\\Shanghai 200240, China}
\affiliation[d]{Shanghai Jiao Tong University Sichuan Research Institute,\\Chengdu 610213, China}
\affiliation[e]{ Jinping Deep Underground Frontier Science and Dark Matter Key Laboratory of Sichuan Province, \\Liangshan 615000, China}
\affiliation[f]{Vacree Technologies Co., Ltd., \\Hefei 230094, China}
\affiliation[g]{School of Chemical Engineering, Zhengzhou University,\\Zhengzhou 450042, China}
\affiliation[h]{School of Physics, Sun Yat-Sen University, \\Guangzhou 510000, China}
\affiliation[i]{Institrte of Refrigeraion and Cryogenic, Shanghai Jiao Tong University,\\Shanghai 200240, China}
\affiliation[j]{Department of Physics, University of Maryland, College Park, \\Maryland 20742, USA}
\affiliation[k]{Yalong River Hydropower Development Company, Ltd., \\Chengdu 610051, China}

\emailAdd{wangzhou0303@sjtu.edu.cn}
\emailAdd{hongloumeng@sjtu.edu.cn}
\emailAdd{shaobo.wang@sjtu.edu.cn}

\abstract{The PandaX-xT cryogenic distillation system for removing radon (Rn) is designed to operate at a high liquid flow rate of 856 kg/h, equal to 5 liters per minute (LPM), to meet the required Rn concentration. However, it is challenging to maintain a continuous supply of cooling power (20 kW) for the Rn-removal system. This work proposes a throttling refrigeration system to address the power supply problem for the next-generation PandaX-xT dark matter detection experiment. The system utilizes carbon tetrafluoride (R14) refrigerant to recover power based on heat pump principles. The cooling power from the liquid xenon (LXe) in the reboiler at 178 K is transferred to liquefy the gaseous xenon (GXe) in the cryostat through the phase change circulation of R14, enabling xenon (Xe) to return to the detector in the same state as when it was extracted. Consequently, a stable circulation for liquid distillation is established. Given the high cost of Xe, a research and development (R\&D) experiment has been conducted to validate the feasibility of this high-cooling-recovery system by using ethanol as a substitute for Xe. Experimental results indicate that the proposed system could achieve a cooling power recovery of 25 kW, with a recovery efficiency of 99$\%$±0.5$\%$. Additionally, process simulation and optimization of the throttling refrigeration cycle are performed using Aspen Hysys. The deviation between the simulation and experimental results is found to be <3.64$\%$. These findings demonstrate that implementing Rn-removal distillation at a high flow rate is feasible, eliminating the need for liquid nitrogen or cryocoolers. This system could save approximately 3650 $m^3$ of liquid nitrogen per year or reduce power consumption by 230 kW.}


\arxivnumber{} 

\begin{document}
\maketitle
\flushbottom

\section{Introduction}
\label{sec:intro}

Xenon (Xe) is an important medium for direct dark matter detection experiments because of its unique physical properties: Liquid Xenon (LXe) has a high density of approximately $\sim$3 g/cm³ which enables it to effectively resist external $\gamma$ and $\beta$ rays through a self-shielding mechanism. Simultaneously, its low energy threshold at electron volt (eV) level, high energy resolution and excellent ionization yield coupled with a scintillation light yield of over 30 photons / keV make it highly sensitive to the faint signals from Weakly Interacting Massive Particles (WIMPs)~\cite{1,2,3,4}. To enhance the sensitivity of dark matter detection, large-scale experiments using Xe are being developed around the world. Notable examples include the use of dual-phase xenon time projection chamber (TPC) technology by XENON, LUX-ZEPLIN (LZ) and PandaX Collaborations, which have let to significant advancements in the field ~\cite{5,6,7}. However, these detectors poses challenges, particularly the need to suppress radon (Rn) contamination~\cite{8}. WIMPs can elastically scatter off xenon nuclei and deposit energy, producing primary scintillation and ionization signals in the LXe, such interactions are referred to as nuclear recoils (NRs). Electronic recoils (ERs), induced by incident gamma rays or electrons, deposit energy onto the atomic electrons of xenon and yield the two signals in a different ratio. Owing to the detector response, ERs constitute a significant background to WIMP searches. The radon isotope $^{222}\mathrm{Rn}$ and its daughter isotope $^{214}\mathrm{Bi}$ with $\beta$ -emitter can generate energetic electrons that deposit energy in the LXe through electromagnetic interactions. This process generates fake signals and degrades the sensitivity of the detector by influencing the background of ERs in LXe ~\cite{9}. $^{222}\mathrm{Rn}$ is continuously produced within the detector materials from the decay of $^{238}\mathrm{U}$ and infiltrates the LXe~\cite{10}, as the detectors scale up in size, the amount of released Rn also rises due to the larger surface area of the materials used. 

\begin{figure}[htbp]
	\centering
	\includegraphics[width=1\textwidth]{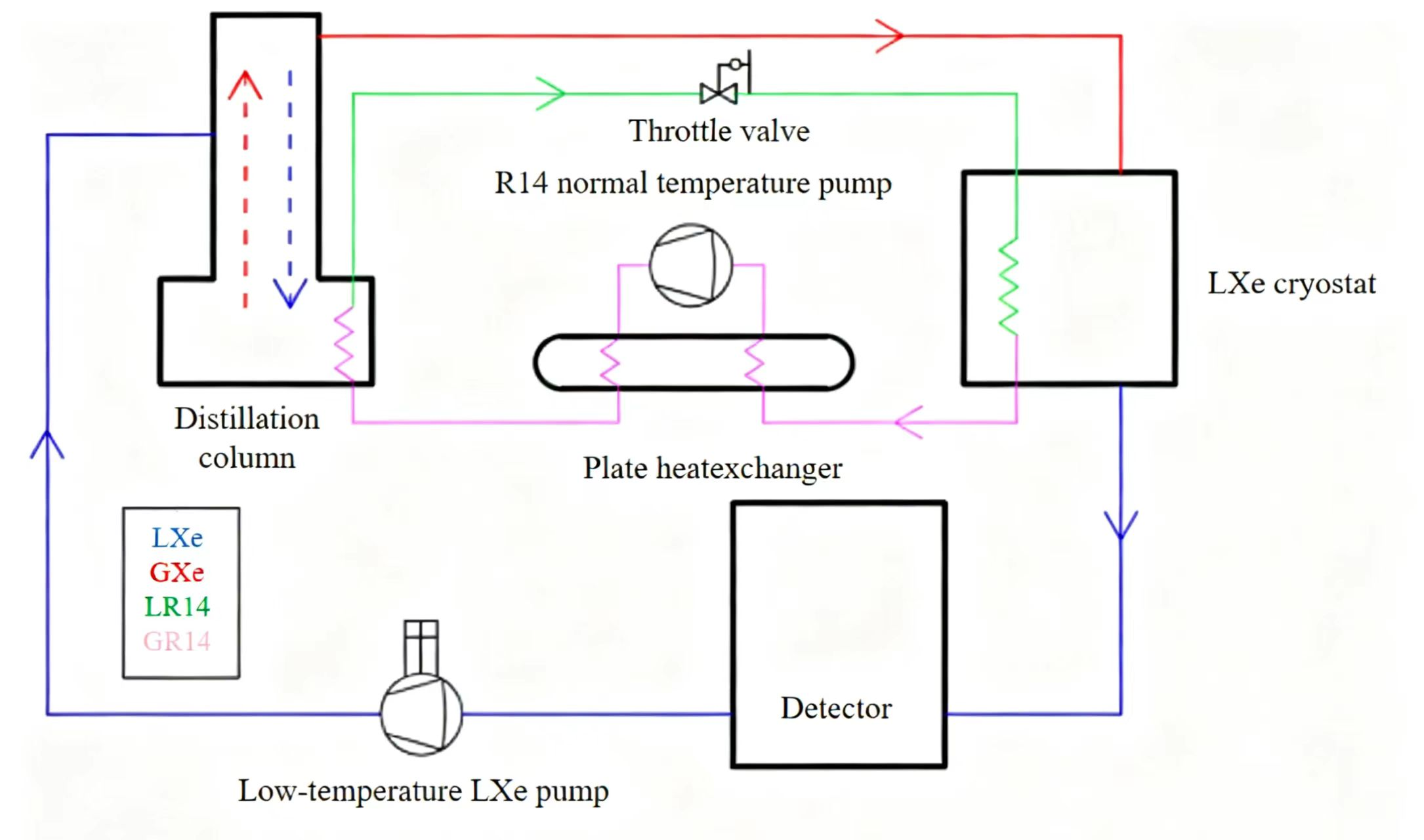}
	\caption{\label{fig1}Diagram of the Rn-removal distillation with Xe loop and R14 loop } 	
\end{figure}

PandaX-xT is the next-generation of Chinese large-scale LXe dark matter detection experiment, which requires the Rn concentration to be limited to 0.5 $\mu$Bq/kg (equivalent to 5×$10^{-26}$ mol/mol)~\cite{9}. This necessitates the development of online cryogenic distillation, which operates alongside the dark matter detector as illustrated in Figure~\ref{fig1}, Regulating the distillation process is regarded as an efficient method for eliminating Rn besides the strict material selection and processing techniques~\cite{11}. Research indicates that increasing the online distillation flow rate is vital for suppressing the Rn background~\cite{12}. In PandaX-xT, LXe online distillation circulation is designed to achieve a flow rate of 856 kg/h (5 LPM) to meet the Rn concentration requirement of 0.5 $\mu$Bq/kg, because of the much higher density of LXe (2854 $kg/m^3$) compare to GXe (18.8 $kg/m^3$) at the operation condition of 178~K and 2$\times10^5$ Pa. The stable operation of the LXe online distillation relies on two critical processes as shown in Figure~\ref{fig1}: Firstly, the LXe extracted from the detector is pumped into the distillation tower, where a bottom reboiler continuously provides heating power to vaporize the LXe. The Xe vapor from the reboiler enters the distillation column, undergoing vapor-liquid exchange as well as heat and mass transfer with the refluxed LXe condensed by the top condenser at the packings. This process effectively separates Xe and Rn. Given that the boiling points of Rn and Xe at atmospheric pressure are 211 K and 165 K, respectively, the purified Xe gas with low Rn concentration is extracted from the condenser at the top of the distillation tower and flows into the LXe cryostat. Secondly, the LXe cryostat acts as a buffer, requiring sufficient cooling power to liquefy the purified Xe gas. Once liquefied, the LXe flows back to the detector at a temperature and pressure similar to those in the detector to avoid disruption of its condition, thus completing the LXe circulation process.

The online distillation circulation for LXe have to address the challenges of thermodynamic equilibrium present in the system. Traditional refrigeration schemes, such as cryocoolers and liquid nitrogen cooling, exhibit limitations as the scale of experiments and the flow rates for distillation increase. In the Rn removal distillation system of PandaX-xT, the reboiler requires a heating power of 22~kW to vaporize LXe. Two Gifford-McMabon (GM) cryocoolers (AL600) installed at the condenser provide 2 kW of cooling power for refluxing, which means the cryostat needs to supply at least 20 kW of cooling power to condense the remaining gaseous xenon (GXe). To meet this 20~kW cooling requirement, 20 commercial cryocoolers, each providing 1 kW of cooling power at 178 K and consuming 11.5 kW of electricity (AL600, Cryomech. Inc), have to operate in parallel. This leads to a total power consumption exceeding 230 kW. Alternatively, utilizing the traditional liquid nitrogen cooling method to attain 20~kW of cooling power requires at least 362~kg (about 450~L) of liquid nitrogen every hour, based on its latent heat of 200 kJ/kg. This amounts to a daily liquid nitrogen consumption exceeding 10 $m^3$, which far exceeds the supply capacities of the laboratory. Consequently, developing controllable cooling power recovery technology is imperative. XENON collaboration has developed a Xe circulation refrigeration system for online Rn distillation that integrates low-radioactivity GXe compressors and efficient heat exchangers~\cite{13}. This system has significantly curtailed liquid nitrogen consumption by 67$\%$, reducing it from 1300 kg/day to 430 kg/day, while supporting a distillation flow rate of 67 kg/h~\cite{13}. Despite these advancements, the system still partially relies on liquid nitrogen cooling and encounters limitations in obtaining higher distillation flow rates, primarily due to constraints with the low-radioactivity GXe compressors. To meet the high-flow rate requirement of 856 kg/h for online distillation circulation while minimizing reliance on liquid nitrogen and cryocoolers, and ensure long-term reliable system operation, it is essential to develop a large cooling power recovery system for the PandaX-xT Rn removal distillation system.

This work proposes an innovative carbon tetrafluoride (R14)-based throttling refrigeration system to facilitate cooling power transition between the reboiler of the distillation tower and the LXe cryostat. This system utilizes the latent heat of R14 during phase changes, along with temperature and pressure regulation. Based on the throttling refrigeration technology, the PandaX-xT large cooling power recovery system could provide 20 kW of cooling power for the Rn-removal cryogenic distillation column, making large flow rate online distillation feasible. This paper outlines the design principles and experimental methodology of the high cooling power recovery throttling refrigeration system for the PandaX-xT cryogenic distillation system in Section~\ref{sec2}. Experimental results and analyses are presented in Section~\ref{sec3}. Section~\ref{sec4} covers the simulation and optimization of this system using Aspen Hysys, followed by conclusions in Section~\ref{sec5}.

\section{System Design and Experimental Methodology}
\label{sec2}
\subsection{Principle Design of the High Cooling Power Recovery Throttling Refrigeration System}
The throttling refrigeration cycle is a dynamic process that leverages the phase change of the refrigerant to provide cooling power~\cite{14,15}. In this cycle, the compressed high-pressure, high-temperature gas refrigerant is condensed into a liquid in the condenser. It then undergoes adiabatic expansion through a throttling device, leading to a sharp pressure drop and partial flash vaporization, which produces a low-temperature two-phase flow. The refrigerant subsequently enters the evaporator, where it absorbs heat (thus providing cooling power) through vaporization before returning to the compressor to complete the circulation. The principle example is illustrated in Figure~\ref{fig1}, our throttling refrigeration system for high cooling power recovery in PandaX-xT utilizes refrigerant R14, whose boiling point is similar to Xe, the latent heat of R14 contributes to the cooling power recovery. In the reboiler, R14 is pressurized to higher pressure of 1.5 MPa by a compressor, raising its boiling temperature to 199.05 K, which is higher than the boiling point of the LXe in the reboiler at 178 K with 0.2 MPa pressure. Consequently, the gaseous R14 is condensed and provides heating power to vaporize the LXe. In the LXe cryostat, the boiling point of R14 at 0.4 MPa is 168.08 K after throttling, which is lower than that of GXe in it (178 K at 0.2 MPa). As a result, the liquid R14 vaporizes, absorbing heat and providing cooling power to the cryostat, in order to facilitate the condensation of the GXe. 

\begin{figure}[htbp]
	\centering
	\includegraphics[width=1\textwidth]{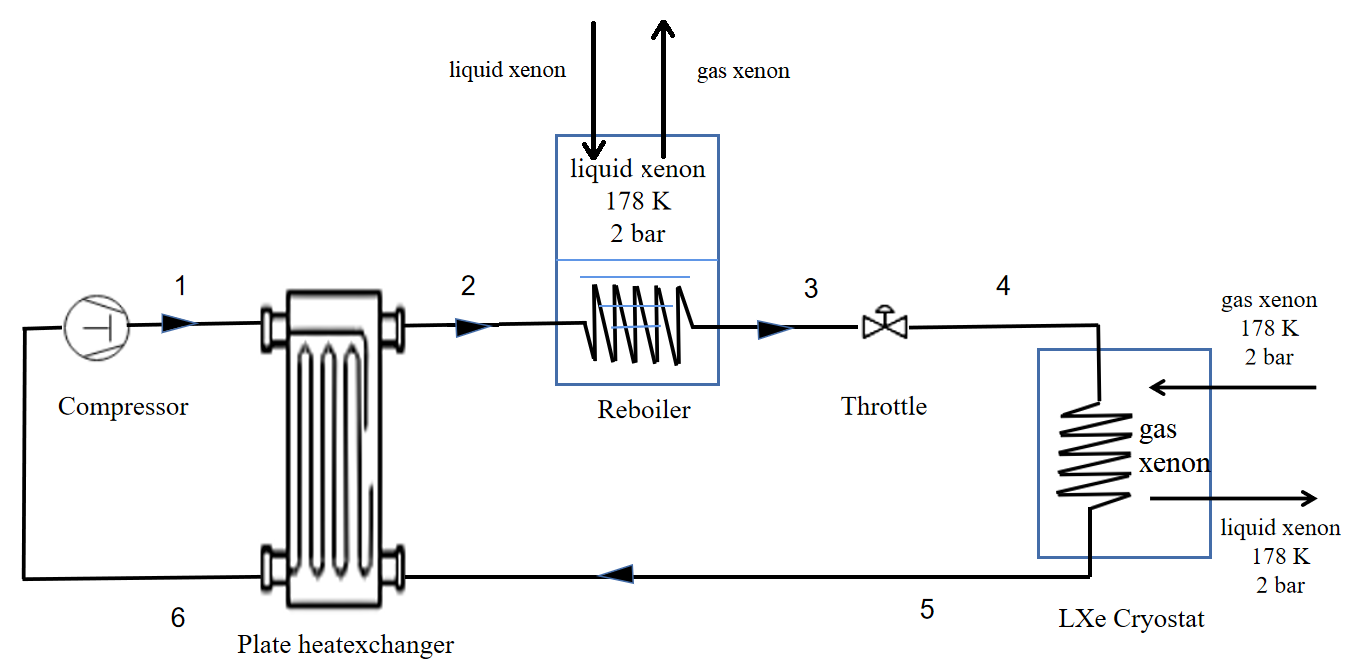}
	\caption{\label{fig2}The flow diagram of the R14 high cooling power recovery throttling refrigeration system. Point 1: compressor outlet and plate heat exchanger inlet; Point 2: plate heat exchanger outlet and reboiler inlet; Point 3: reboiler outlet and throttle valve inlet; Point 4: throttle valve outlet and LXe cryostat inlet; Point 5: LXe cryostat outlet and plate heat exchanger inlet; Point 6: plate heat exchanger outlet and compressor inlet.}  	
\end{figure}

\begin{figure}[htbp]
	\centering
	\includegraphics[width=1\textwidth]{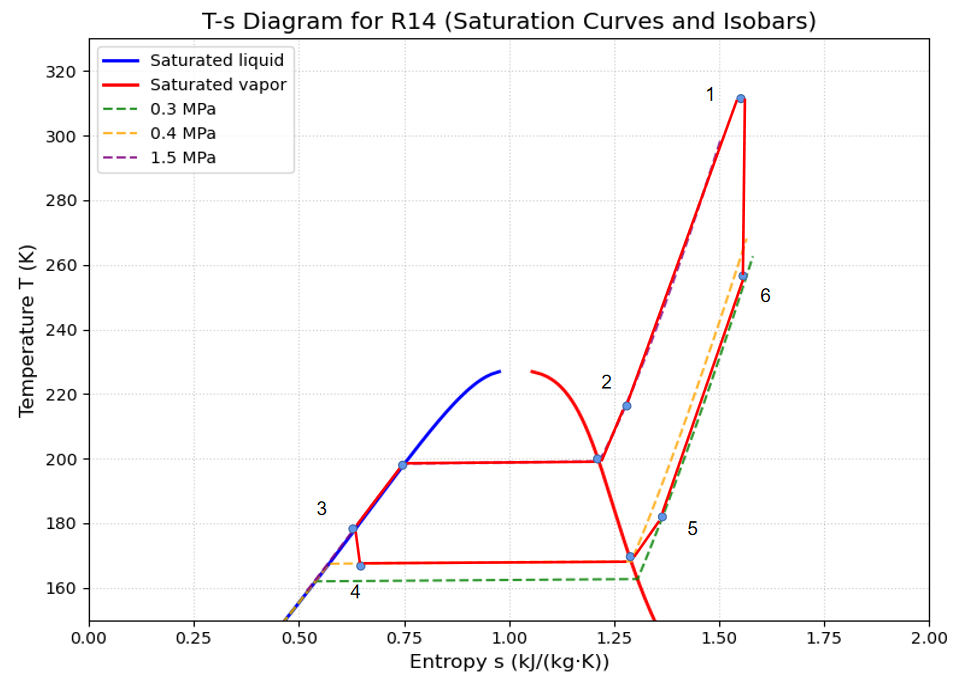}
	\caption{\label{fig3}Temperature-Entropy (T-S) diagram of the high cooling power recovery throttling refrigeration system} 	
\end{figure}

The designed process scheme of the high cooling power recovery throttling refrigeration system is shown in Figure~\ref{fig2}, where the operational conditions are determined through theoretical calculations with ideal assumption, causing differences to the actual experimental data shown in Section 3. Figure~\ref{fig3} presents the corresponding Temperature-Entropy (T-S) diagram~\cite{16,17}. In the figures, from point 6 to point 1, the pressure of the gaseous R14 is isentropic compressed~\cite{18} from 0.3 MPa to 1.5 MPa by the compressor, with the temperature increasing from  276 K to 310 K. It then flows into a plate heat exchanger where it is pre-cooled by the low-temperature R14 from the LXe cryostat (point 5), thereby enhancing cooling power recovery efficiency defined by Equation \ref{eq:efficiency} and preventing liquid slugging towards the compressor, and its temperature drops to 210 K at point 2 before entering the reboiler. In the reboiler, the gaseous R14 contacts the LXe at 178 K through the stainless steel coil and undergoes isobaric condensation from point 2 to point 3, transitioning to the sub-cooled liquid R14 of 180 K, while releasing heat to the LXe in the reboiler. After point 3, the liquid R14 then flows through the throttle valve, where adiabatic expansion occurs due to a sudden reduction in flow cross-section~\cite{19}. Here, the pressure drops sharply from 1.5 MPa to 0.4 MPa at point 4, falling below its saturation pressure, which triggers flash evaporation. A portion of the liquid R14 vaporizes instantaneously, forming a gas-liquid R14 mixture at 168 K. This process is approximately isenthalpic process with an irreversible increase in entropy. After throttling from point 3 to point 4, the gas-liquid R14 at 168 K and 0.4 MPa enters the LXe cryostat via the stainless steel coil, where it contacts the GXe at 178 K, This interaction allows R14 to evaporate continuously by absorbing 20 kW of heating power from the Xe condensation, ultimately transforming into superheated gas at 176 K at point 5. Point 6 denotes where the low-temperature gaseous R14 is heated to 276 K via the plate heat exchanger before returning to the compressor, thereby completing the thermodynamic cycle. The cooling power recovery circulation is achieved by matching the pressure gradient and phase changing temperature zones, as well as efficient coupling of R14 condensation and heat release in the reboiler with its evaporation and heat absorption in the LXe cryostat.

\subsection{ R\&D Experimental Setup the refrigeration system}

\begin{figure}[ht]
	\centering
	\includegraphics[width=1\textwidth]{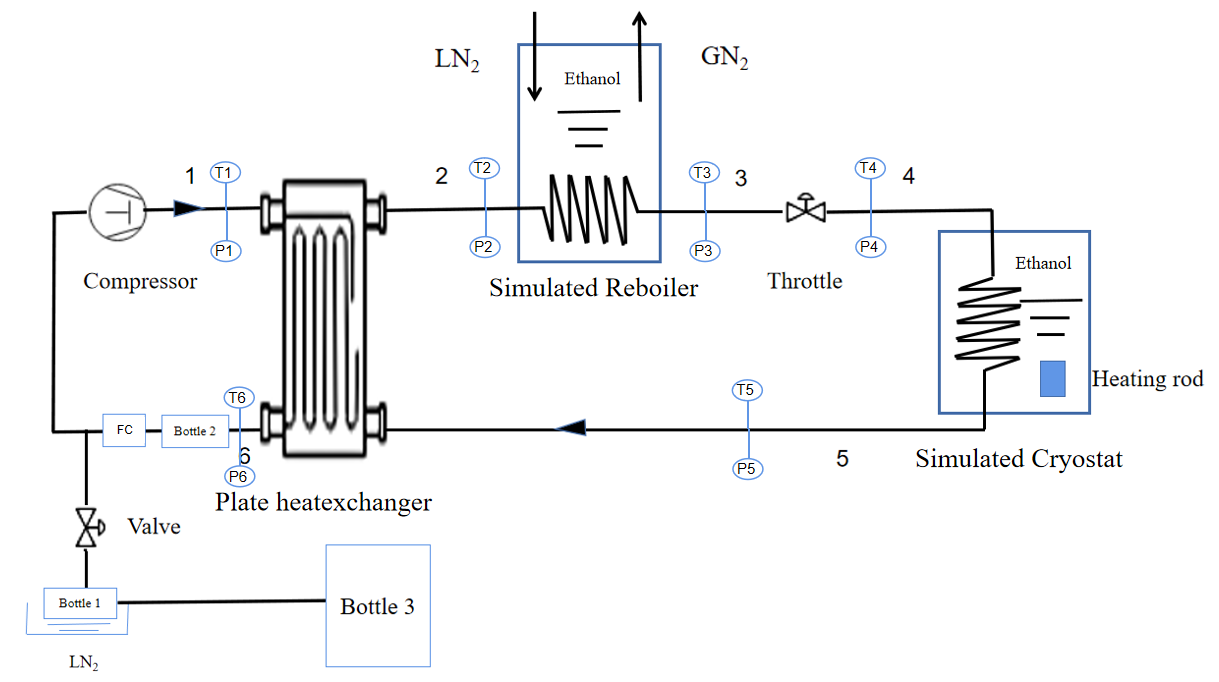}
	\caption{\label{fig4}R\&D experimental flow diagram of the high cooling power recovery throttling refrigeration. In which T is the temperature measuring point and P is the pressure measuring point. Point 1:compressor outlet; Point 2: simulated reboiler inlet; Point 3: simulated reboiler outlet; Point 4: simulated cryostat inlet; Point 5: simulated cryostat outlet; Point 6: compressor inlet. } 		
\end{figure}

To verify the feasibility of the high cooling power recovery system for the PandaX-xT Rn-removal cryogenic distillation, a R\&D experimental setup was designed, configured and implemented. Ethanol was used as a substitute for Xe due to the high cost of the latter. The block diagram representing the R\&D experimental configuration is depicted in Figure~\ref{fig4}. The simulated reboiler and cryostat were filled with ethanol to replicate the actual operating conditions. Liquid nitrogen was continuously supplied to the simulated reboiler via the copper coil to maintain the ethanol temperature at 178 K, thus simulating the behavior of LXe in the reboiler, where the R14 condenses and releases heat. In the simulated cryostat, ethanol replaced the GXe, where liquid R14 evaporated and absorbed heat. To prevent a continuous descent in temperature, ten heaters of 2.5 kW each were installed to quantify the cooling power generated by the evaporation of liquid R14 from the coil heat exchanger in the simulated cryostat: the cooling power was evaluated by the heating power of the heaters and the temperature of the ethanol. The increase, decrease and stabilization of the temperature indicated the heating power of the heaters is above, below and equal to the cooling power, which also defined as the cooling power capacity of the system.

Compare to the R\&D experiment, the R14 coil heat exchanger has higher heat transfer coefficient in the actual PandaX-xT Rn-removal distillation system in principle, because of the xenon phase-change out of the coil heat exchanger in the reboiler and the cryostat, where the heat transfer contains the evaporation and condensation heat exchanges, respectively.

The heat transfer coefficient of the Xe evaporation out of the coil heat exchanger in the reboiler is calculated using Mostinski Equations as below ~\cite{20}:
\begin{equation}
h_{reboiler} = 0.106 \cdot P_c^{0.69} \cdot q^{2/3} \cdot F_p
\label{eq2-1}
\end{equation}

\begin{equation}
F_p = 1.8(p_r)^{0.17} + 4p_r^{1.2} + 10p_r^{10}
\label{eq2-3}
\end{equation}

\begin{equation}
p_r = P / P_c 
\label{eq2-2}
\end{equation}

Where $h_{reboiler}$ is the heat transfer coefficient outside the tube in the LXe of the reboiler at 178K and 2$\times10^5$ Pa (2~atm), W/(m²·K); $q$ is the heat flux, which is 1.96 kW/m² when heating power is 20 kW with the tube of outer diameter of 25.4 mm and length of 128 m; $P_c$ denotes the critical pressure of 5.842 MPa, and $P$ is the operation pressure. The Mostinski equation is a classic semi-empirical correlation based on the principle of equilibrium state, which is primarily employed for estimating the heat transfer coefficient in nucleate boiling regions for non‑polar pure fluids, under moderate‑to‑low pressure conditions (with reduced pressure $p_r$ significantly below 1). The liquid xenon in the reboiler status of 178 K and 2 atm ($p_r\approx 0.035$) corresponds to the low‑temperature and low‑pressure condition, so the equation is applicable.

The heat transfer coefficient for the GXe condensation out of the coil heat exchanger in the cryostat is calculated using Nusselt Equation as below ~\cite{21}:

\begin{equation}
h_{cryostat} = 0.729 \left[ \frac{(\rho_l - \rho_v) ^2g \cdot h_{fg} \cdot k_l^3}{\mu_l \cdot d \cdot \Delta T} \right]^{1/4}
\label{eq2-4}
\end{equation}

Where $h_{cryostat}$ is the heat transfer coefficient outside the tube in the GXe of the cryostat at 178 K and 2 atm, W/(m²·K); $\rho_l$ represents the LXe density with a value of 2854.7 $kg/m^3$; $\rho_v$ denotes the Xe vapor density, 18.7 $kg/m^3$; $g$ is the gravity, 9.8 $m/s^2$; $h_{fg}$ is the latent heat equal to 93.7 kJ/kg; $k_l$ signifies the heat conductivity coefficient of 0.068 $W/(m\cdot K)$; $\mu_l$ is the Xe dynamic viscosity, $4.4\times 10^{-4} Pa\cdot s$; $d$ is the outer diameter of the tube, 25.4 mm; and $\Delta T$ is the temperature difference between the inner and outer tube. Under the low reduced pressure ($p_r\approx 0.035$), the classical Nusselt film condensation theory is applicable for estimating the heat transfer coefficient of the xenon vapor condensation.

During the R\&D experiment, the heat transfer between the ethanol and the R14 coil heat exchanger occurred via natural convection, resulting in a lower heat transfer coefficient.

The heat transfer coefficient for natural convection of the ethanol out of the coil heat exchanger in both the simulated reboiler and the simulated cryostat could be calculated using the classic equation as follows~\cite{21}:

\begin{equation}
h = Nu_f\cdot k_l'/d'
\label{eq2-5}
\end{equation}

\begin{equation}
Nu_f = 0.48(GrPr)^{1/4}(Pr_f-Pr_w)^{0.11}
\label{eq2-6}
\end{equation}

\begin{equation}
Gr = \frac{g \beta \Delta T d'^3}{\nu^2}
\label{eq2-7}
\end{equation}

Where $h$ is heat transfer coefficient out of the tube in the ethanol of the simulated reboiler or simulated cryostat at 178 K and 1$\times10^5$ Pa, W/(m²·K); $Nu_f$ denotes the Nusselt number; $Pr$ represents the Prandtl number, which is 38.3 for ethanol; $Pr_f$ and $Pr_w$ is the Prandtl numbers corresponding to the fluid temperature and wall temperature, respectively; $Gr$ is Grashof number; $\beta$ is coefficient of volume expansion, 0.001 $K^{-1}$; $\Delta T$ is the temperature difference between the wall and the fluid; $d'$ is the outer diameter of the tube, 25.4 mm, and $\nu$ is the kinematic viscosity, $3.8\times 10^{-6} m^2/s$. This classic equation is applicable to the outer surface of horizontal circular tube in a large liquid space with laminar natural convection (Rayleigh number $Ra\le10^{9}$), which is suitable for our experimental conditions with the ethanol Ra of $1.93\times 10^{6}$.

According to the calculations, the overall heat transfer coefficient of the heat exchanger in the simulated reboiler is 153.9 W/(m²·K) (Equation~\ref{eq2-5} ), while the actual reboiler recorded a coefficient of 364.8 W/(m²·K) (Equation~\ref{eq2-1} ). Similarly, the overall heat transfer coefficient for the simulated cryostat is calculated to be 142.1 W/(m²·K) (Equation~\ref{eq2-5} ), compared to 1347.2 W/(m²·K) in the actual cryostat (Equation~\ref{eq2-4} ). These results indicate that the comprehensive heat transfer performance in the actual distillation system is significantly superior to that of the current R\&D experiment. Therefore, the ethanol-based $R\&D$ experiment for simulation serves as a reliable and effective equivalent testing method. If the coil heat exchanger used in the $R\&D$ experiment could achieve the expected heat transfer performance, its effectiveness under actual phase-change conditions would be validated. This helps to confirm the reliability of the high cooling power recovery throttling refrigeration design for the PandaX-xT cryogenic distillation system. The investigation could also verify whether the key structural parameters of the coil heat exchanger, such as the tube diameter, parallel arrangement and the total length, meet the heat transfer requirements of the system. Furthermore, the analysis provides insight into the flow resistance characteristics under realistic operating conditions, guiding the structural design of the PandaX-xT Rn-removal cryogenic distillation system and the implementation of the high cooling power recovery system.

\begin{figure}[H]
	\centering
	\includegraphics[width=1\textwidth]{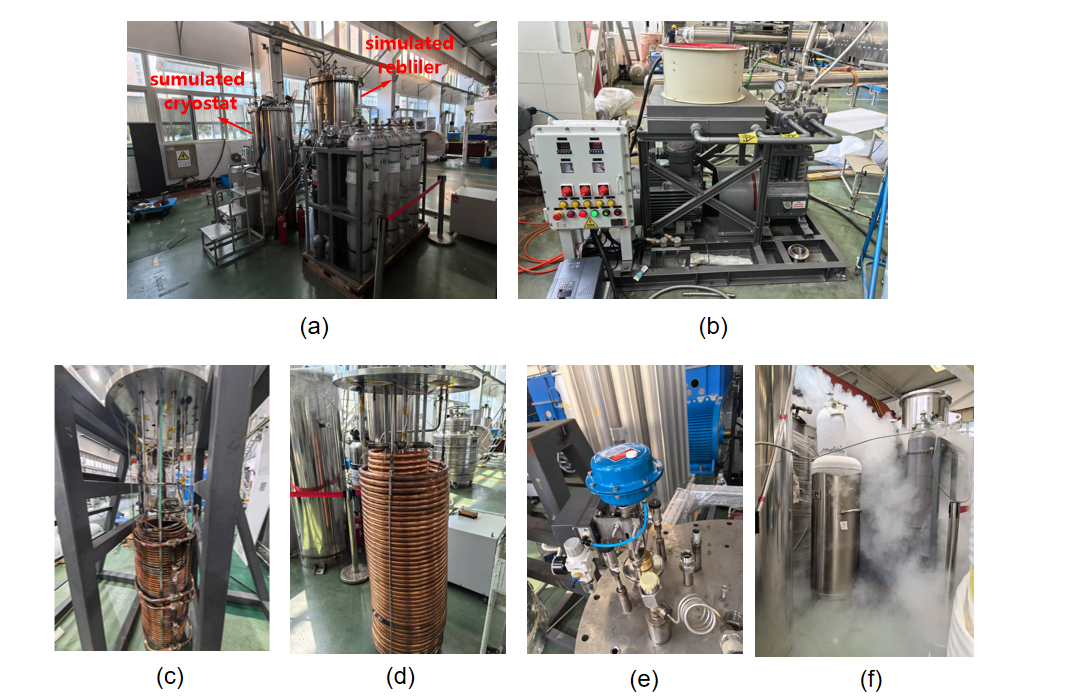}
	\caption{\label{fig5}(a) R\&D experimental setup of the high cooling power recovery throttling refrigeration system  (b) Compressor (c) Simulated reboiler heat exchanger and plate heat exchanger (d) Simulated cryostat heat exchanger (e) Throttle valve (f) R14 recuperation system using liquid nitrogen} 	
\end{figure}

The experimental setup, illustrated in Figure~\ref{fig5}, consists of five main components: compressor, plate heat exchanger, coil heat exchanger in simulated reboiler, coil heat exchanger in simulated cryostat, and throttle valve. The inlet pressure range of the piston compressor (Bengbu Aiot. co., Model ZTW-0.65/4-25, 18.5~kW, air-cooling) is 0.1 to 0.5 MPa, with an outlet pressure range of 1 to 2.5 MPa and a maximum flow rate of 0.26 kg/s. The plate heat exchanger (SWEP, Model B80Hx36/1P), which has an efficiency of 80\% as determined from experimental data~\cite{22}, is employed to exchange the sensible heat between the high-pressure hot gaseous R14 from the compressor and the low-pressure cold R14 from the simulated cryostat for pre-cooling, while preventing liquid impact toward the compressor. The pneumatic proportional valve (Jiangsu Cryote Co., BMFA-HG-600) is used as throttle valve for precise control.

As portrayed in Figure~\ref{fig5}(b), After compression in the compressor (1600 mm length $\times$ 1000 mm width $\times$ 1300 mm height), the high-temperature, high-pressure gaseous R14 enters a plate heat exchanger for sensible heat exchange with the returning low-temperature R14 vapor from the simulated cryostat. The pre-cooled gas then flows into a simulated reboiler for condensation as shown in Figure~\ref{fig5}(a)(inner diameter Ø600 mm, height 1950 mm), the reboiler contains a submerged copper coil (128 m, outer diameter Ø25.4 mm, wall thickness 1.5 mm) in an ethanol bath of about 380~kg, as shown in Figure~\ref{fig5}(c), with temperature measurement points at both the top and bottom. The heat exchanger utilizes a four-pass parallel configuration to accommodate high flow rates and reduce flow resistance. The condensed high-pressure liquid R14 then undergoes isenthalpic expansion through a throttle valve (Figure~\ref{fig5}(e), inner diameter Ø19.1 mm, length 990 mm), transitioning to a low-temperature, low-pressure two-phase state. The valve features a pneumatic proportional actuator integrated with the Programmable Logic Controller (PLC) to enable remote and high-precision flow control. After the throttle valve, the two-phase mixture R14 enters the simulated cryostat with a inner diameter of 550 mm and height of 1637 mm as illustrated in Figure~\ref{fig5}(a), whose heat exchanger comprises two parallel copper coils (134 m, outer diameter Ø25.4 mm, wall thickness 1.5 mm) as shown in Figure~\ref{fig5}(d). The coils are immersed in an ethanol bath (about 280~kg) with upper and lower temperature measurement points. To apply a controllable heat load, ten heating rods (each rated at 2.5 kW) are uniformly distributed throughout the ethanol of the simulated cryostat. As R14 vaporizes, it absorbs heat as it flows through the coils. The vaporized low-temperature R14 then returns to the plate heat exchanger for reheating before re-entering the compressor, thus completing the cycle. By regulating the compressor, throttle valve and heating power, the thermodynamic performance and heat transfer characteristics of the refrigeration system could be systematically evaluated across various operating conditions.

\section{Experimental Results and Analysis}
\label{sec3}
\subsection{Experimental results of the system}

During the pre-cooling process, the throttle valve was fully opened, and liquid nitrogen at a rate of 3.77 LPM was introduced into the liquid nitrogen coil in the simulated reboiler to pre-cool the ethanol to below 200 K.

After the precooling process, the ethanol temperature in both the simulated reboiler and simulated cryostat was brought down to approximately 190 K. At this point, the compressor was activated and high-cooling power recovery circulation began, the compressor was adjusted using a frequency converter, and the throttle valve opening was adjusted synchronously. Through the coordinated control of these two components, the mass flow rate of R14 in the system cycle was stabilized at approximately 0.2 kg/s with the compressor frequency of 50 Hz and throttle valve opening of 80\%.
Eight electric heating rods immersed in the ethanol of the simulated cryostat were sequentially turned on, providing a total stabilized input power of 20.2 kW, to simulate the heating power released from the product GXe during condensation in the LXe cryostat of the PandaX-xT Rn-removal cryogenic distillation system.

\begin{figure}[H]
	\centering
	\includegraphics[width=0.8\textwidth]{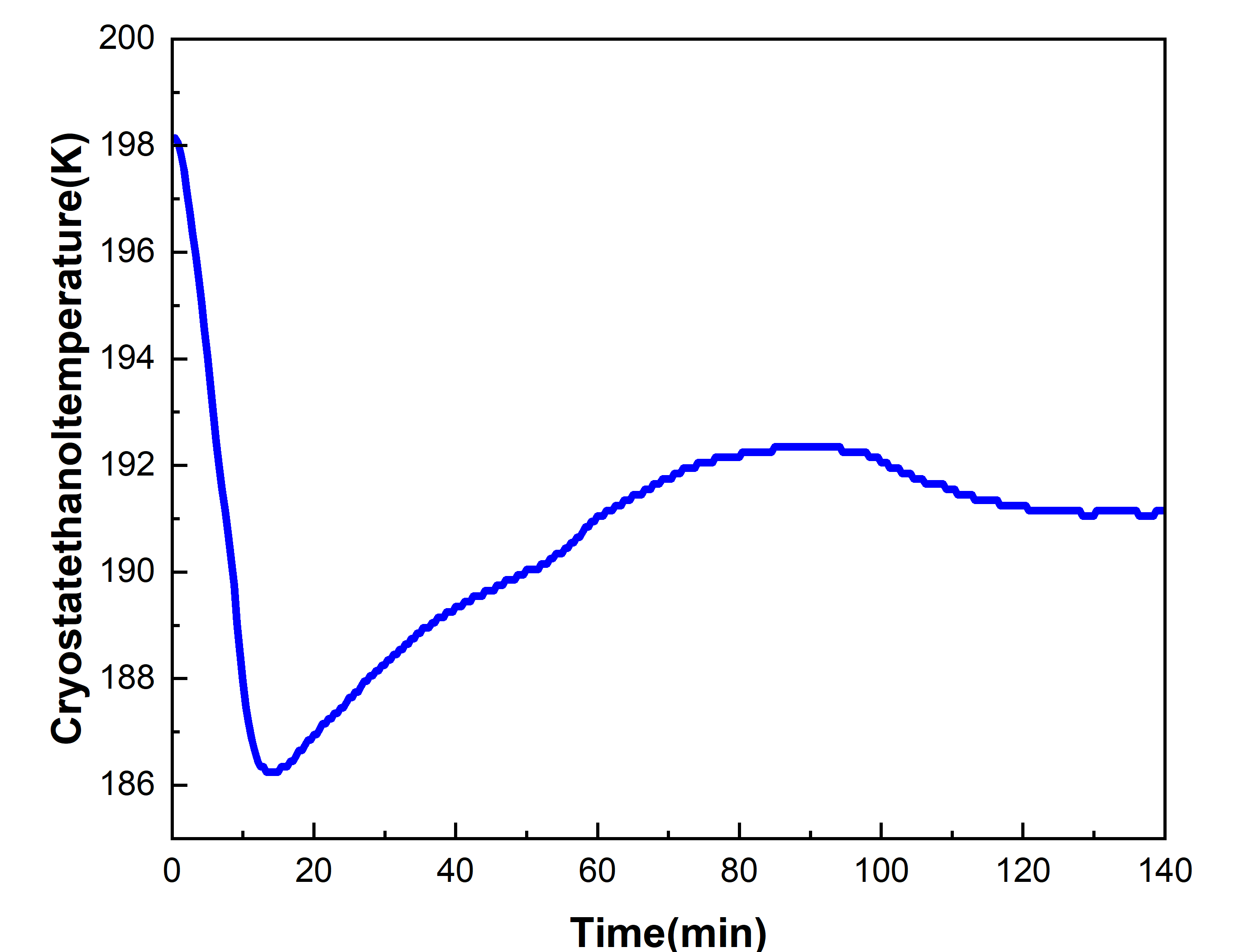}
	\caption{\label{fig6-2}Ethanol temperature meatsured against the time in the simulated cryostat} 	
\end{figure}

Figure~\ref{fig6-2} illustrates the thermal load response of the system to this step. At the initial stage of the heating, after 10 min's delayed response of the system when the heating power of the electric rods withstand the heat capacity of the ethanol in the simulated cryostat, the instantaneous input thermal power caused the ethanol temperature to rise over the course of approximately 70 minutes, with an average heating rate of about 0.37 K·min$^{-1}$. Around the 90 minute mark, the temperature reached a local peak of 192~K. Subsequently, The cooling power generated by the R14 evaporation in the coil heat exchanger gradually increased, as the heat transfer temperature difference increasing enhanced the performance of the heat exchanger. Consequently, this cooling power surpassed the constant heating power supplied by the electric heating rods, causing the temperature to start declining slowly at an average rate of about 0.18 K·min$^{-1}$ till stabilize at 191~K($\pm$0.1~K).

\begin{figure}[H]
	\centering
	\includegraphics[width=0.8\textwidth]{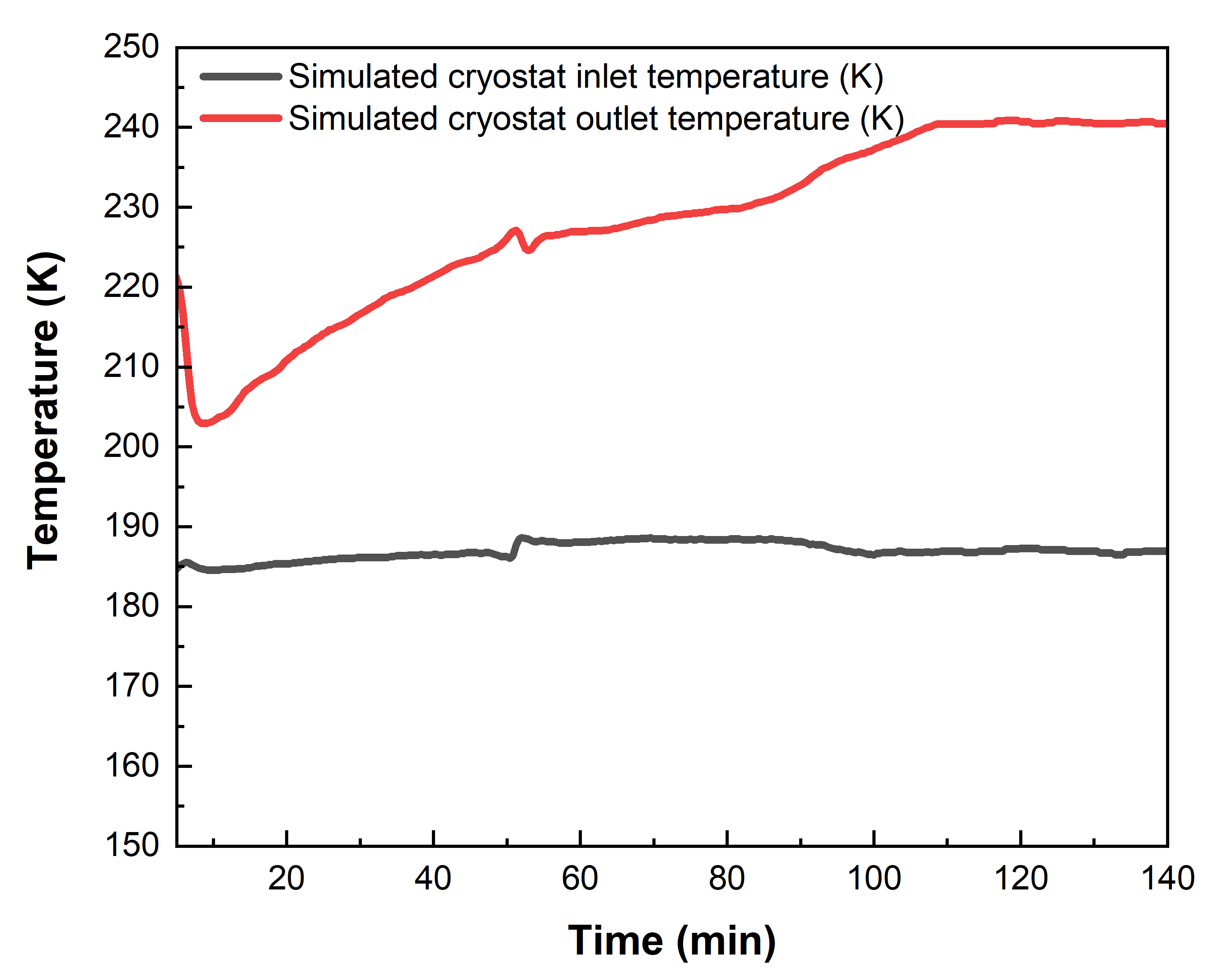}
	\caption{\label{fig7-2}The inlet and outlet temperatures of the simulated cryostat} 	
\end{figure}

As shown in Figure~\ref{fig7-2}, after the introduction of a 20.2~kW external electric heating load, the system began to establish a dynamic thermal equilibrium. A notable observation during this process was the temperature variation of the R14 working fluid at both the outlet and the inlet of the simulated cryostat. The outlet temperature rose from approximately 202~K at the beginning of heating and eventually fluctuated around 240~K ($\pm$0.5~K) under quasi-steady conditions. This temperature increase of about 40~K indicates that the R14 fluid, after transitioning from two-phase flow to saturated vapor in the coil heat exchanger, continued to absorb heat and became superheated vapor. The increase in outlet superheat implies that the total heat absorbed per unit mass of the R14 has risen, with the increment attributed to the elevation of its gaseous sensible heat. Thus, under these operating conditions, the refrigeration mechanism operates as a combination of "latent heat refrigeration" and "sensible heat refrigeration", with the latter providing additional cooling power. 

In contrast to the significant change in outlet temperature, the inlet temperature of the R14 fluid at the simulated cryostat remained stable throughout the process, controlled within a narrow range of 185~K to 188~K. This stability signifies that the R14 condensation process in the upstream simulated reboiler functions smoothly, delivering a stable high-pressure liquid to the throttle valve. A stable inlet state ensures that the sensible heat of the superheated gaseous R14 at the outlet of the simulated cryostat could be effectively absorbed by the simulated reboiler through the plate heat exchanger, thus preventing thermodynamic deteriorate and maintaining stability in the entire circulation system.

\begin{figure}[htbp]
	\centering
	\includegraphics[width=0.8\textwidth]{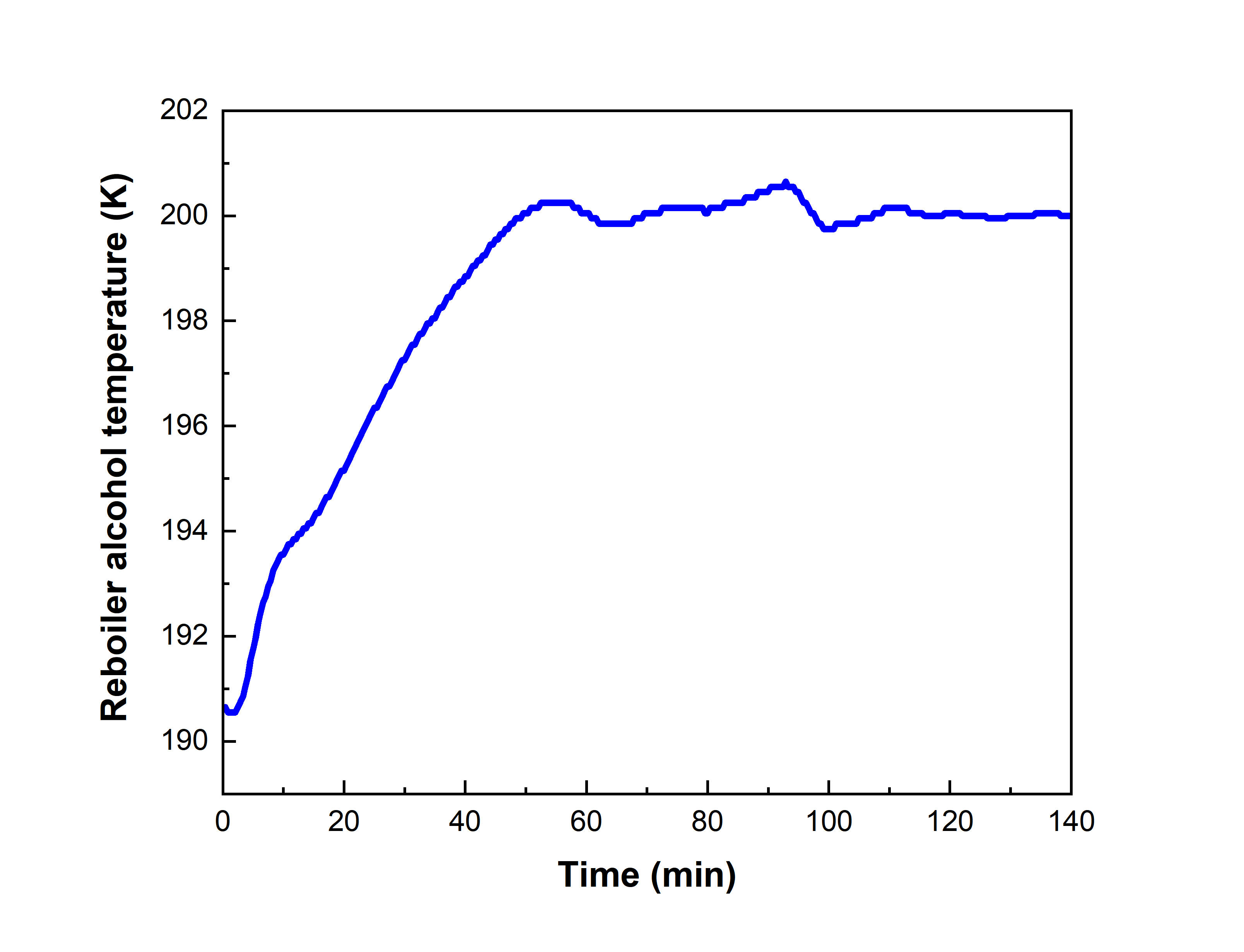}
	\caption{\label{fig8-2}The ethanol temperature evolution in the simulated reboiler}	
\end{figure}

After applying an external heat load and allowing the system to reach a steady state, the variation in ethanol temperature within the simulated reboiler is illustrated in Figure~\ref{fig8-2}. Starting from an initial temperature of approximately 190.5~K, the ethanol temperature gradually increased before stabilizing at around 200~K ($\pm$1~K), with minor fluctuations occurring during subsequent long-term operation. As the superheated R14 gas discharged from the compressor underwent precooling and condensation within the coil heat exchanger, it released substantial latent heat along with some sensible heat, which elevated the ethanol temperature. Concurrently, the liquid nitrogen coil immersed in the ethanol acted as an external cold source, continuously extracting heat and mitigating the temperature rise. During the initial stage, the heat released by R14 condensation likely surpassed the cooling capacity of the liquid nitrogen coil heat exchanger, resulting in an increase in ethanol bath temperature. As the ethanol temperature rose, the growing temperature difference between the ethanol and the liquid nitrogen enhanced the heat transfer rate of the liquid nitrogen coil, thereby augmenting the cooling supply. Once the cooling power of the liquid nitrogen coil heat exchanger balanced the condensation heat load of R14, the ethanol temperature reached a dynamic equilibrium and remained stable, ultimately settling at 200~K ($\pm$1~K), as exemplified in Figure~\ref{fig8-2}.

\begin{figure}[H]
	\centering
	\includegraphics[width=0.8\textwidth]{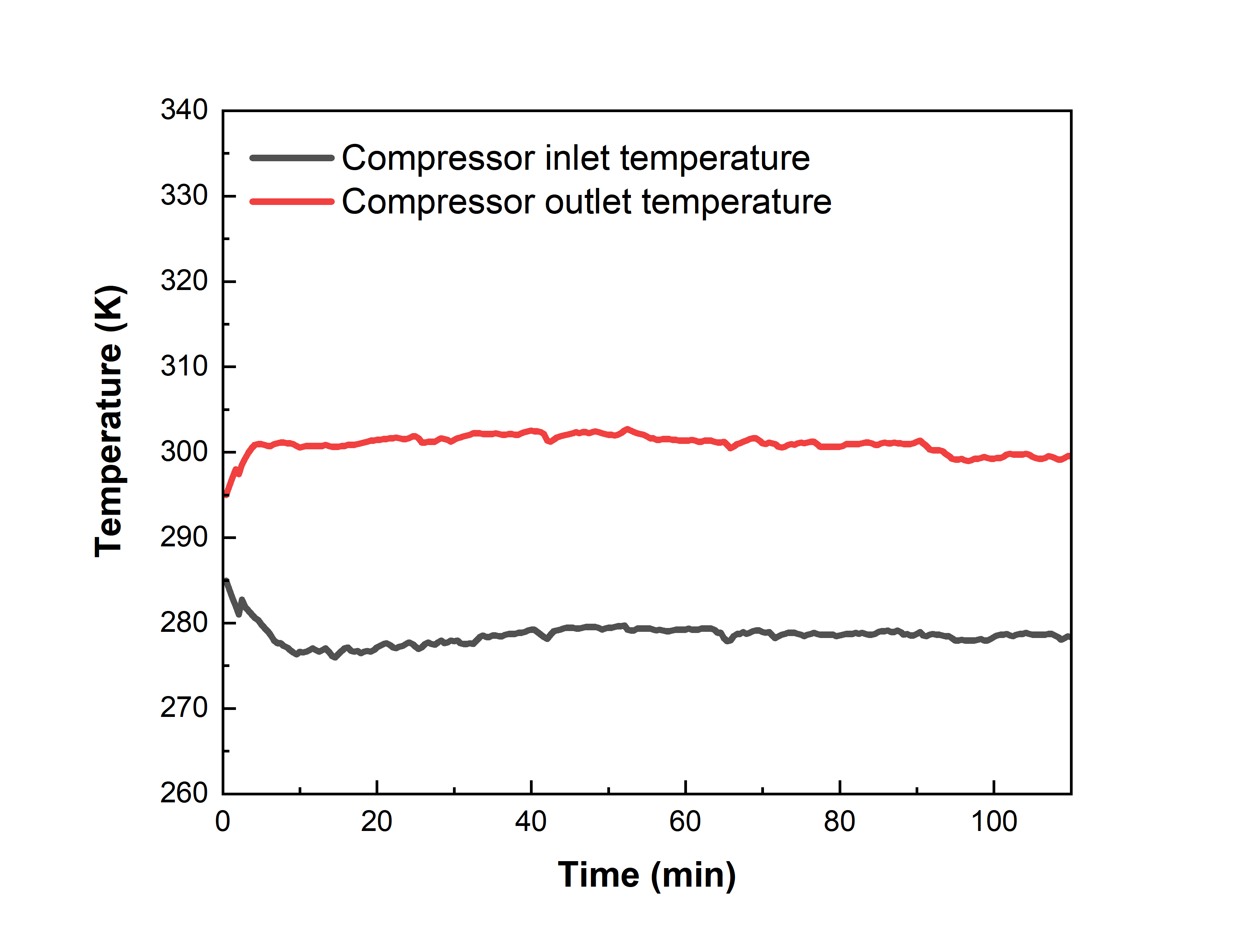}
	\caption{\label{fig9-2}The inlet and outlet temperatures of the compressor}	
\end{figure}

The inlet and outlet temperatures of the compressor are vital indicators reflecting the thermodynamic state and regulation effectiveness of the system, as depicted in Figure~\ref{fig9-2}. At the beginning of the system startup, the compressor inlet temperature was about 285~K. As the external thermal load was applied and the system gradually approached a new equilibrium, the inlet temperature progressively decreased and eventually stabilized at around 276~K ($\pm$1~K). A stable and relatively low suction temperature is beneficial for reducing the discharge temperature and decreasing the suction specific volume, which helps maintain high volumetric efficiency and system mass flow rate.

Initially, the compressor outlet temperature rose from approximately 295~K at startup to around 300~K ($\pm$1~K), followed by a stabilization period. This increase was primarily ascribed to the rising pressure ratio within the system resulting from the increased thermal load and power input during compression. Eventually, the outlet temperature stabilized at around 300~K, indicating that the compressor achieved a preliminary balance between its thermal load and cooling under the new steady-state operating condition.

\subsection{Analysis of the Refrigeration Capacity and Efficiency at Experimental Conditions}

Using the experimentally measured inlet and outlet temperatures of the simulated cryostat when the system was stable, along with their corresponding pressures, the refrigeration capacity of the system could be calculated by the following equation:

\begin{equation}
Q = \dot{m} (h_{5} - h_{4})
\label{eq:refrigeration}
\end{equation}

where $Q$ represents the refrigeration capacity of the system (kW), which is the cooling power recuperated in the simulated cryostat using R14. Here, $\dot{m}$ is the R14 mass flow rate (kg/s),  $h_4$ and $h_5$ are the inlet and outlet enthalpies of R14 in the simulated cryostat evolved by the R14 status along time (kJ/kg), corresponding to point 4 and point 5 in Figure \ref{fig4}, which could be obtained using National Institute of Standards and Technology (NIST) software~\cite{23}. 

\begin{figure}[H]
\centering
\includegraphics[width=0.8\textwidth]{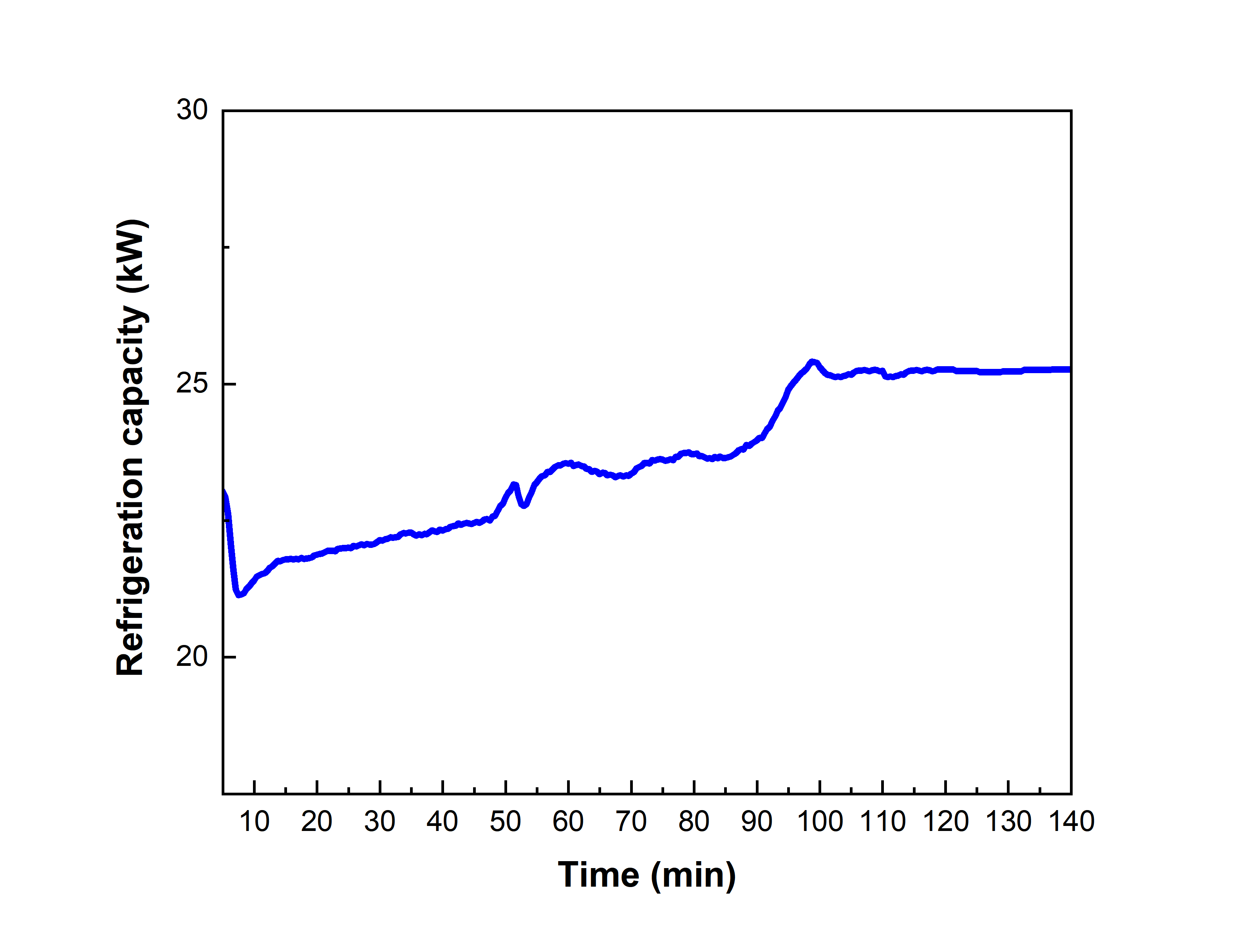}
\caption{Refrigeration capacity of the system}
\label{fig10-2}
\end{figure}

The evolution of the system cooling capacity variation is presented in Figure\ref{fig10-2}. After applying the thermal load using electric heating rods in the simulated cryostat, the cooling capacity increased slowly from an initial value of about 21~kW to a peak of approximately 25.3~kW, before slightly decreasing and fluctuating around 25~kW ($\pm$0.1~kW). This process reflects the dynamic mechanism of the system establishing thermal equilibrium. During the initial stage, the heat transfer temperature difference in the simulated cryostat increased with the rising ethanol temperature, and the R14 mass flow rate changed due to system pressure adjustments, together stimulating an increase in cooling capacity, until dynamic thermal equilibrium was reached.

In the throttling refrigeration system for the recovery of high cooling power, the primary focus is on attaining equilibrium between the heating power released in the reboiler and the cooling power supplied to the cryostat. This balance is essential to avoid the need for additional cooling power from liquid nitrogen or cryocoolers, thereby the system efficiency is the recovery efficiency, defined as Equation \ref{eq:efficiency} :

\begin{equation}
\eta = \frac{Q}{Q_r}
\label{eq:efficiency}
\end{equation}

where $\eta$ is the system efficiency, $Q$ denotes the refrigeration capacity of the system calculated by Equation \ref{eq:refrigeration}, $Q_r$ represents the heat released in the simulated reboiler (kW), which is determined by the enthalpy difference of R14 between the outlet ($h_3$) and inlet ($h_2$) of the simulated reboiler, corresponding to point 3 and point 2 in Figure \ref{fig4}, $Q_r$ could be calculated by  $\dot{m} (h_{2} - h_{3})$. 

\begin{figure}[H]
\centering
\includegraphics[width=0.8\textwidth]{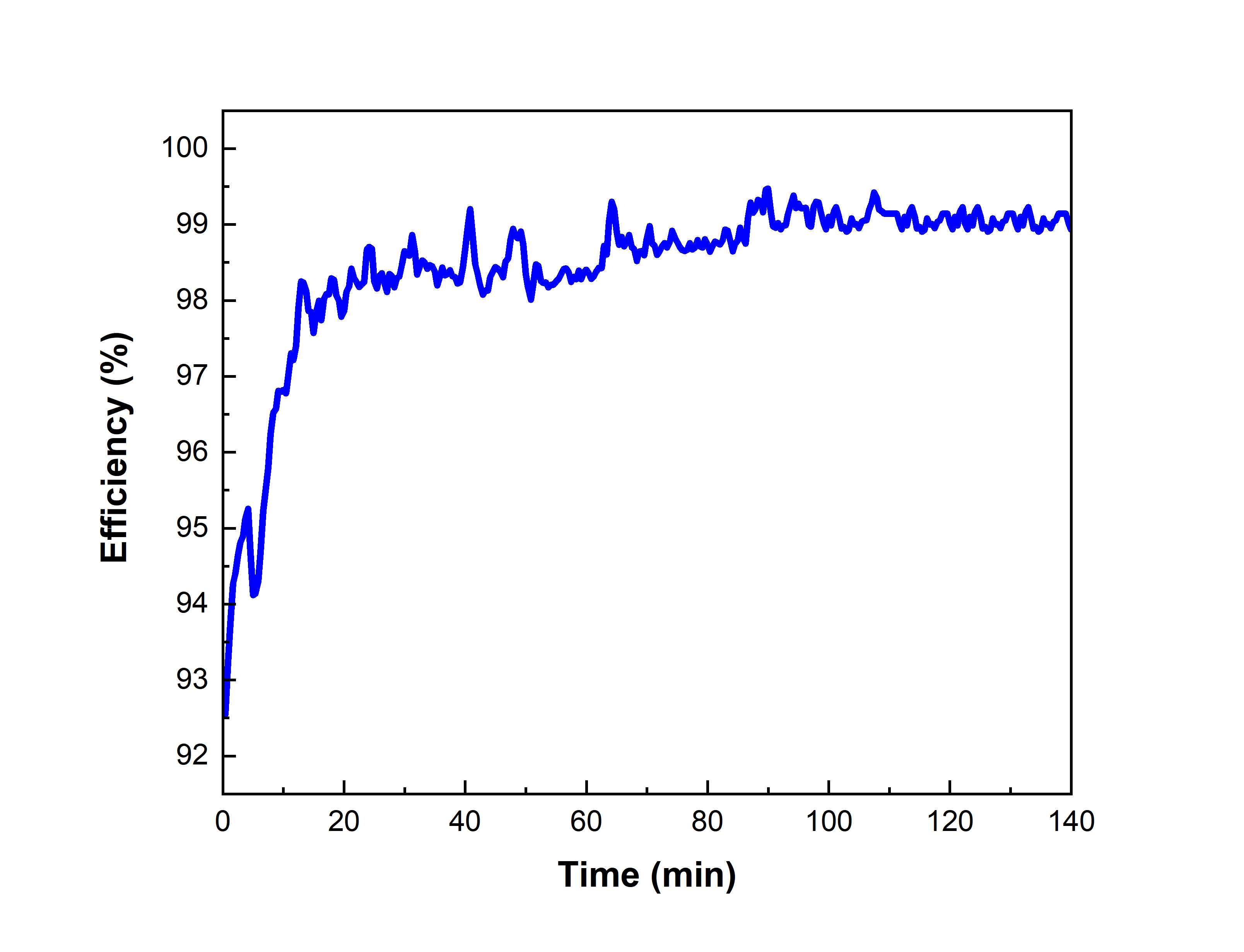}
\caption{System efficiency plotted against the time}
\label{fig11-2}
\end{figure}

The system efficiency calculated from the experimental data is shown in Figure \ref{fig11-2}. Following system startup, the efficiency gradually improved from an initial value of about 92.5\% to 99\% $\pm$ 0.5\%. This implies that the heat released by the condensation of R14 in the reboiler and the cooling absorbed by the  evaporation of R14 in the cryostat were numerically approaching equivalence. This balance is crucial for the stable operation of the PandaX-xT Rn-removal cryogenic distillation system, as it ensures that the heating power on the reboiler side is well matched with the cooling power on the cryostat side. This balance prevents system overpressure caused by excessive Xe vaporization and instability in liquid levels due to excessive Xe condensation.

\section{Simulation and Optimization of the System}
\label{sec4}
\subsection{Process Simulation Model}
Aspen Hysys, a process simulation software , is utilized to construct the steady-state process model of the high cooling power recovery throttling refrigeration system, and the Peng-Robinson (PR) equation is employed for the thermodynamic calculations~\cite{24}:
\begin{equation}
P = \frac { R T } { V - b } - \frac { a \alpha ( T ) } { V ( V + b ) + b ( V - b ) }
\end{equation}

\begin{equation}
\alpha ( T )=[1+\kappa(1-\sqrt\frac{T}{T_c})]^2
\end{equation}

where $P$ is the pressure, $a$ is the molecular attraction parameter;  $V$ is the molecular volume; $b$ is the molecular volume parameter, both calculated according to the critical temperature $T_{c}$ and critical pressure $P_{c}$ of the component. Additionally, $\alpha(T)$ is a temperature-dependent correction function that incorporates the acentric factor $\omega$ ($\kappa = 0.37464 + 1.54226\omega - 0.26992\omega^{2}$). Specifically, the critical properties for $\mathrm{R14}$ ($\mathrm{CF}_{4}$), such as $T_{c} = 227.5\,\mathrm{K}$, $P_{c} = 3.74\,\mathrm{MPa}$, and its acentric factor ($\omega = 0.178$) are entered into the property database.

\begin{figure}[h]
\centering
\includegraphics[width=1\textwidth]{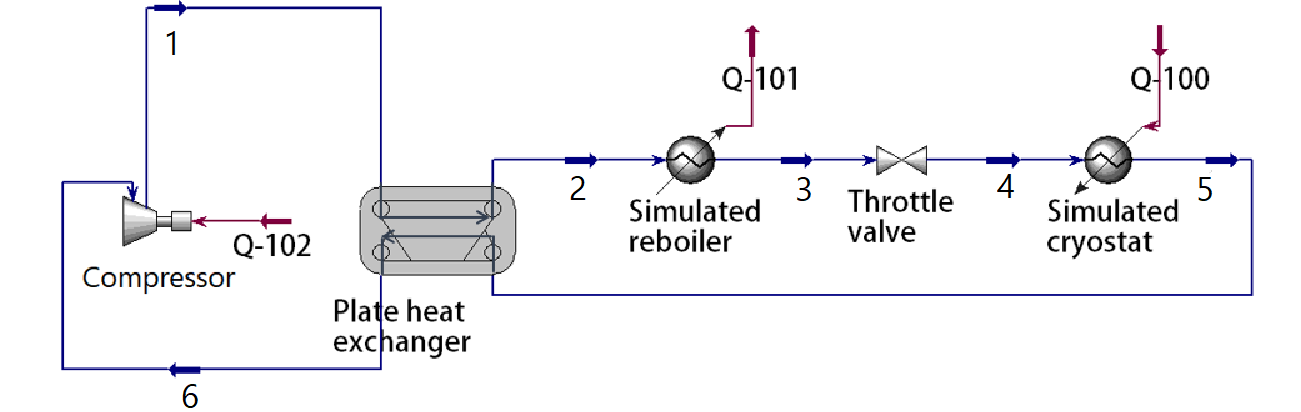}
\caption{Simulation model of the R\&D experiment for the recovery of high cooling power via throttling refrigeration system. }
\label{fig12}
\end{figure}

The model of the throttling refrigeration system, depicted in Figure \ref{fig12}, comprises several key components, including the compressor (modeled as an isentropic process), the plate heat exchanger, the simulated reboiler, the throttle valve, and the simulated cryostat. The air-cooling system of the compressor is thermodynamically simulated using an equivalent cooler module. To validate the reliability of the simulation model, the key boundary conditions are determined based on experimental data collected during stable operation. These include the throttle valve pressure drop of 1.03$\times 10^{6}$ Pa, pre-throttling temperature of 195~K, pre-throttling pressure of 2.47$\times 10^{6}$ Pa, the R14 temperature at the simulated cryostat outlet of 240~K, and the R14 temperature of the compressor inlet of 276~K.
 
\subsection{Comparison of the Simulation Results with the Experimental Outcomes}
From the Aspen Hysys model, whose built-in device parameters (such as heat exchanger efficiency, compressor volumetric efficiency, etc.) are adjusted according to the experimental data to ensure the reliability, the refrigeration capacity of the system is simulated to be 24.09 kW, derived from the temperature and pressure conditions at both ends of the simulated cryostat. The experimental result is 25.0 kW($\pm$0.1~kW), yielding a relative error of 3.64\% between the simulation and experimental refrigeration capacities of the system. The simulation results for the important temperatures at points 2 and 4 are recorded as 193.7 K and 185.8 K, respectively, while the experimental readings are 193.8 K($\pm$0.5~K) and 185 K($\pm$0.5~K). The relative deviations between these temperature results are 0.05\% and 0.43\%, respectively. These findings indicate that the simulation results derived from thermodynamic equilibrium and mass conservation principles, align closely with the experimental data, thereby validating the reliability of the simulation. The minor discrepancies could be attributed to the non-ideal system characteristics and measurement uncertainties, including ignoring the heat leakage from the low-temperature vacuum-insulation pipeline in the ideal simulation model, potential sensor errors during experimentation (e.g., temperature drift, pressure fluctuations), and using simplified assumptions regarding constant isentropic compressor efficiency and fixed heat exchanger pressure drops, which differ from the experimental dynamic behaviors. 

\subsection{Process Optimization}
\subsubsection{Influence of the throttle valve opening on the refrigeration capacity}
Figure\ref{fig13}(a) demonstrates that the pressure drop across the throttle valve influences the temperature after the throttle valve and the mass flow rate of the system. Simulation outcomes from Aspen Hysys indicate that as the pressure drop increases from 0 to 1.8 MPa (with the valve opening decreasing from 100\% to about 40\%), the temperature after the throttle valve declines from 210 K to 180 K, while the mass flow rate reduces from 0.17 kg/s to about 0.03 kg/s. Based on the simulation results shown in Figure\ref{fig13}(a), Figure\ref{fig13}(b) illustrates the impact of the pressure drop of the throttle valve on the heat transfer coefficient and the temperature difference in the simulated cryostat. With the flow rate decreasing from 0.17 kg/s to 0.03 kg/s when the pressure drop increasing from 0 MPa to 1.8 MPa, the heat transfer coefficient of R14 inside the coil heat exchanger correspondingly drops from 142 W/m²·K to about 13 W/m²·K. Conversely, the heat exchanger temperature difference between the R14 and the ethanol increases from 0 K to about 30 K due to the stronger throttling effect. While the reduction in the heat transfer coefficient diminishes the heat exchange efficiency, the increase in the temperature difference tends to improve it. Therefore, an optimal range for the refrigeration capacity could be simulated.What needs to be pointed out is that the R14 amount in the simulation is constant.

\begin{figure}[H]
\centering
\includegraphics[width=0.56\textwidth]{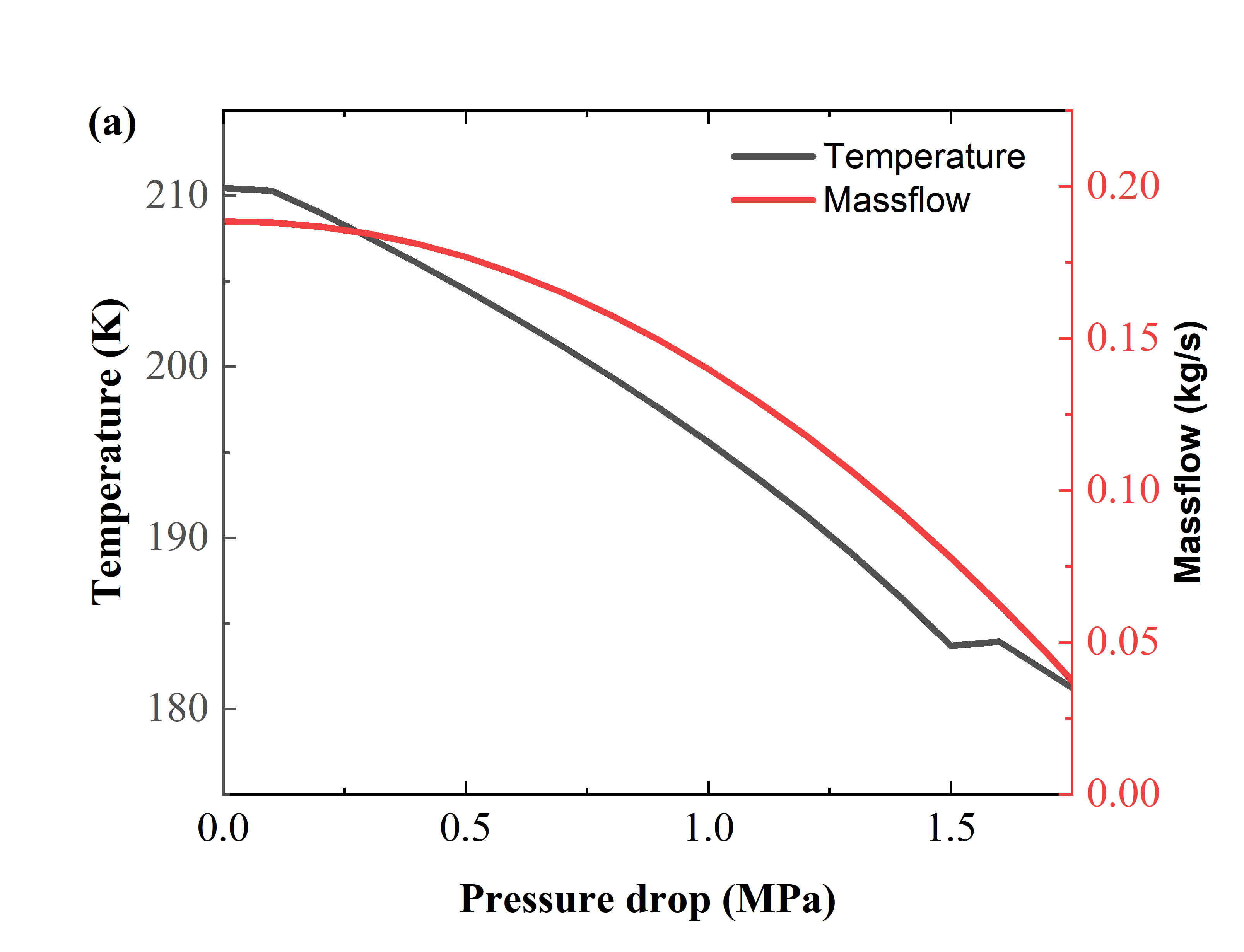}
\end{figure}

\begin{figure}[H]
\centering
\includegraphics[width=0.56\textwidth]{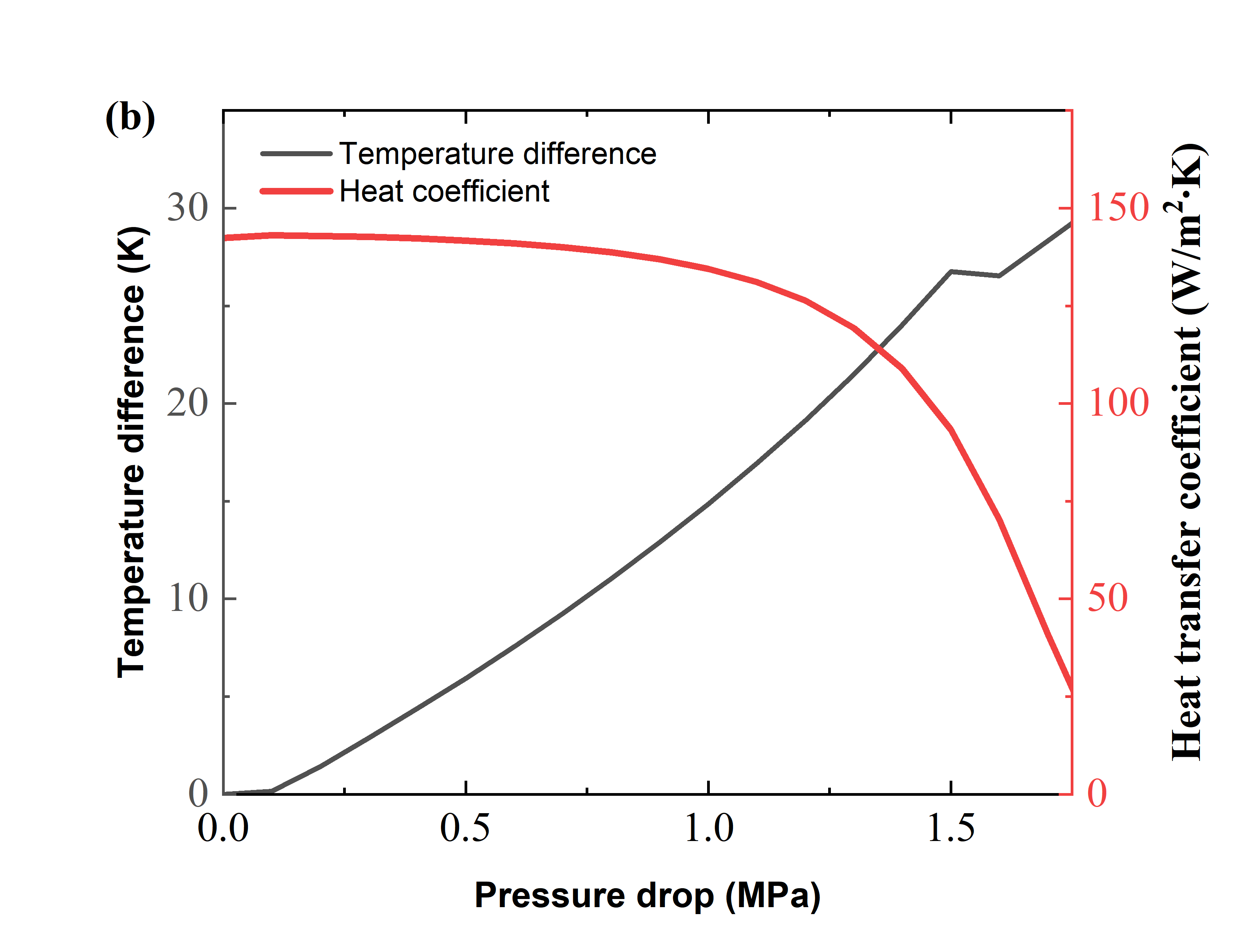}
\vspace{1em}
\caption{(a) The simulation diagram of the flow rate and temperature after the throttle valve influenced by the pressure drop of the throttle valve. (b) The simulation diagram of the heat transfer coefficient and the temperature difference of the heat exchanger in the simulated cryostat influenced by the pressure drop of the throttle valve.}
\label{fig13}
\end{figure}

Based on the calculation and simulation results described above, Figure\ref{fig14}(a) shows the influence of the pressure drop across the throttle valve on the refrigeration capacity of the system. It is observed that the refrigeration capacity initially increases then decreases along the pressure drop rising, peaking at about 28 kW when the pressure drop is 1.4 MPa. Figure\ref{fig14}(b) depicts the relationship between the pressure drop and the valve opening degree. The optimal pressure drop for the throttle valve is found to be 1.4 MPa  (corresponding to a valve opening of 45-50\%), under these conditions, the flow rate is 0.1 kg/s, the heat transfer coefficient is 109 W/m²·K, the temperature after the throttle valve is 167 K, and the heat transfer temperature difference between R14 and the ethanol in the simulated cryostat is 24 K.

\begin{figure}[H]
\centering
\includegraphics[width=0.6\textwidth]{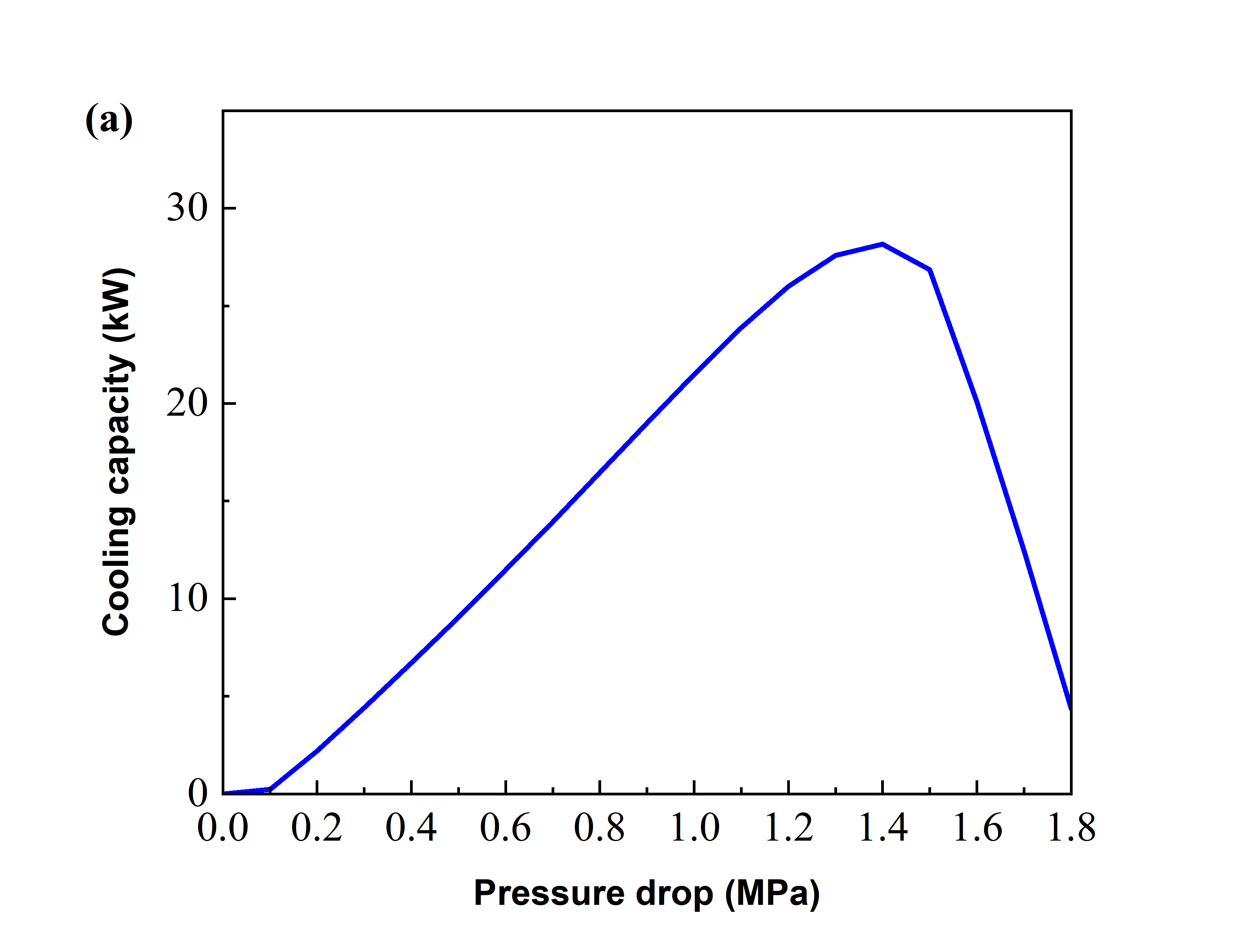}
\end{figure}

\begin{figure}[H]
\centering
\includegraphics[width=0.6\textwidth]{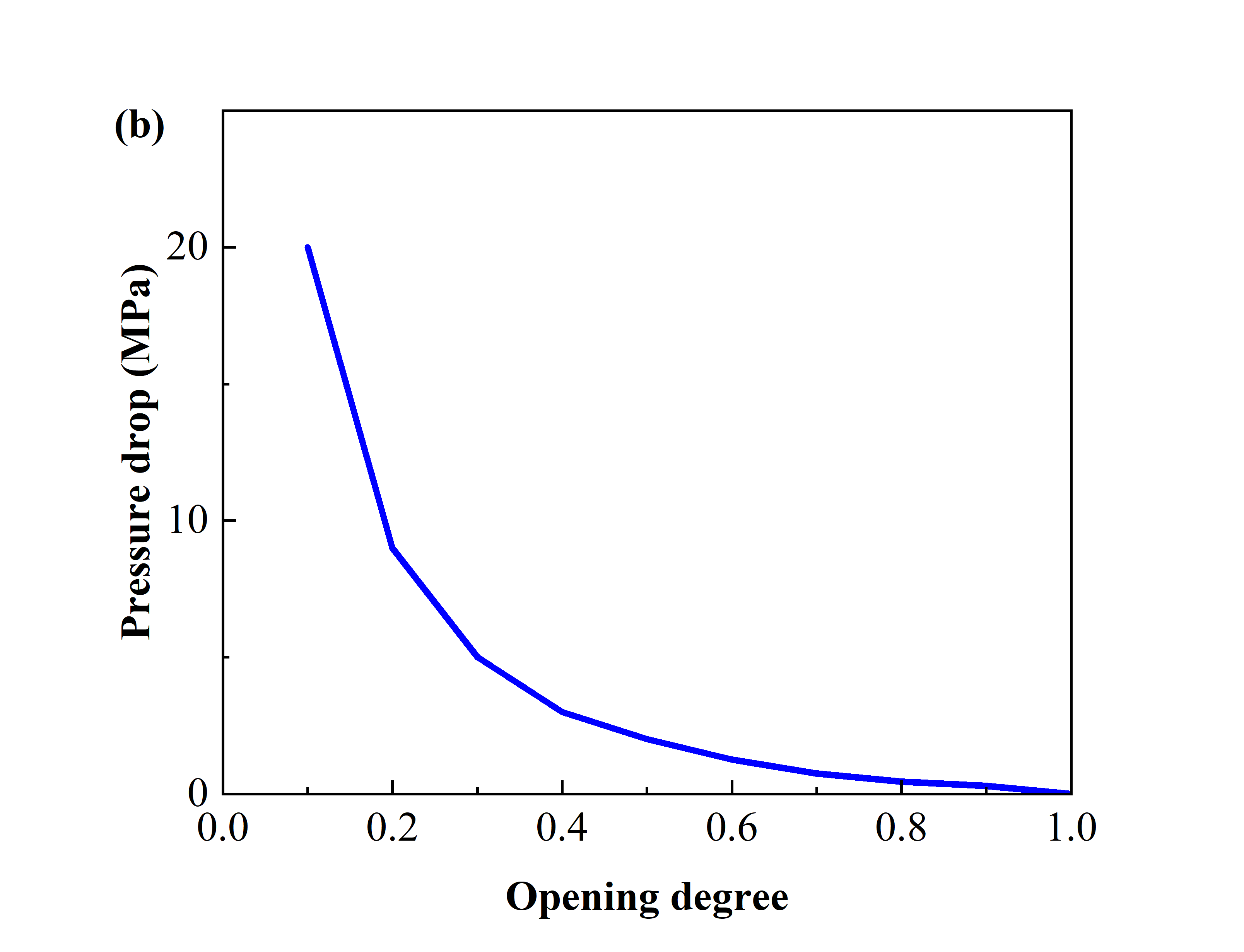}
\vspace{1em}  
\caption{(a) The simulation diagram of the influence of the throttle valve pressure drop to the refrigeration capacity of the system denoted in Equation 3.1. (b) The simulation diagram of the throttle valve pressure drop and its opening degree.}
\label{fig14}
\end{figure}

\subsubsection{Influence of the sub-cooling degree before the throttle valve and the pre-throttling pressure on the refrigeration efficiency}
Figure \ref{fig15}(a) demonstrates the influence of the sub-cooling temperature before the throttle valve to the refrigeration efficiency of the system. The system efficiency is observed to rise from 78\% to 82\% by increasing the sub-cooling temperature from 1 K to 20 K. This improvement could be attributed to a higher sub-cooling degree that reduces the enthalpy of the refrigerant before the throttle valve, which in turn decreases the generation of flash gas after the throttle valve, thus enhancing the specific refrigeration capacity. Figure \ref{fig15}(b) illustrates the positive correlation between pre-throttling pressure and the system efficiency. As the pressure increases from 2.3 MPa to 3.0 MPa, the refrigeration efficiency enhances from 79\% to 85\%. Thermodynamically, the appropriate higher pressures could bolster the refrigeration efficiency.

\begin{figure}[H]
\centering
\includegraphics[width=0.56\textwidth]{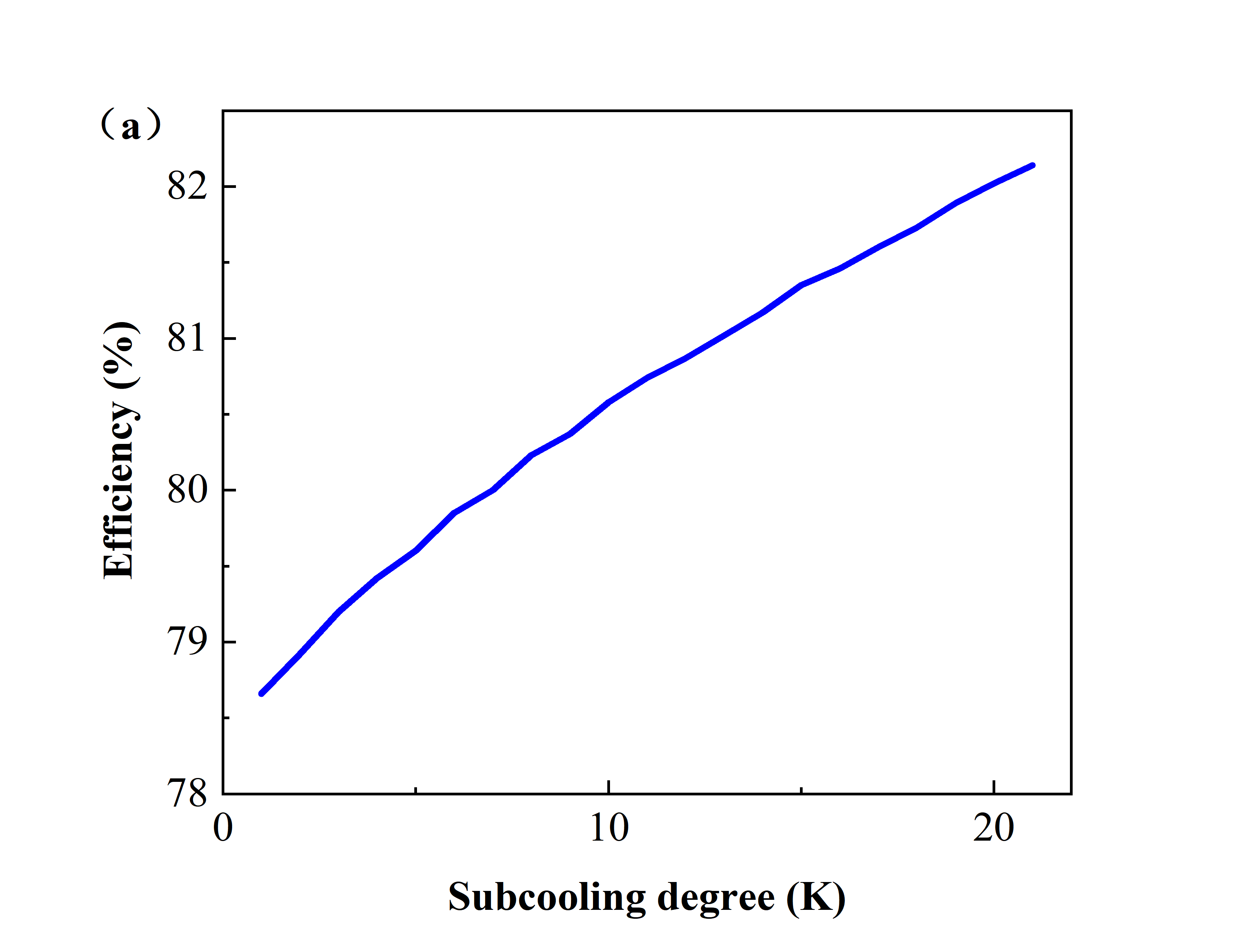}
\end{figure}

\begin{figure}[H]
\centering
\includegraphics[width=0.56\textwidth]{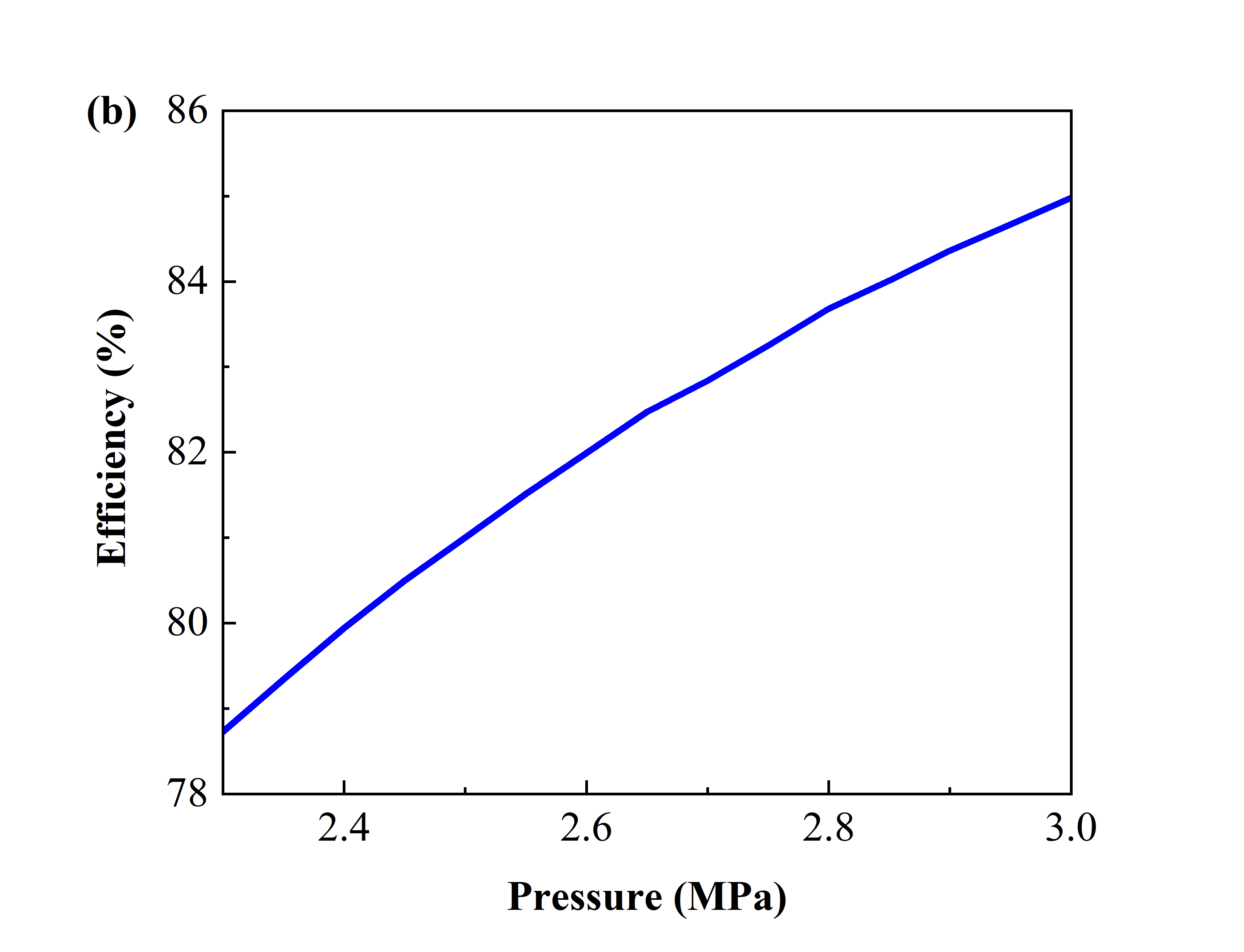}
\vspace{1em}
\caption{(a) Influence of the sub-cooling degree before the throttle valve to the refrigeration efficiency of the system. (b) Influence of the pre-throttling pressure to the refrigeration efficiency of the system.}
\label{fig15}
\end{figure}

\subsubsection{Influence of the R14 temperature of the simulated cryostat outlet and the R14 temperature of the compressor inlet on the refrigeration efficiency}
Figure \ref{fig16}(a) exemplifies the impact of the R14 temperature at the simulated cryostat outlet on the refrigeration efficiency of the system. As the temperature increases from 178 K to 230 K, the refrigeration efficiency boosts from 70\% to 77.5\%. This enhancement is due to higher evaporation temperatures elevating the specific refrigeration capacity. Conversely, Figure \ref{fig16}(b) presents the influence of the R14 temperature at the compressor inlet on the refrigeration efficiency, the efficiency of the system decreases from 83\% to 78\% as the temperature increases from 240 K to 310 K, indicating that a higher R14 inlet temperature of the compressor corresponds to lower refrigeration efficiency. This occurred because the increased temperature of the compressor inlet results in a larger specific heat capacity of the suction gas, leading to a declines of the volumetric efficiency and gas delivery of the compressor. Consequently, power consumption of the compressor increases, diminishing the refrigeration efficiency.

In summary, this study investigates the mechanisms by which various operation parameters influence system performance based on simulation analysis. The optimal throttling pressure drop is 1.4 MPa according to the throttle valve opening of 45-50\%, corresponding to the cooling capacity of 28 kW. Increasing both the sub-cooling before throttling and pre-throttling pressure could improve system efficiency. Appropriately increasing the cryostat outlet temperature is also beneficial for improving efficiency, while lowering compressor suction temperature can further enhance efficiency. 

\begin{figure}[H]
\centering
\includegraphics[width=0.6\textwidth]{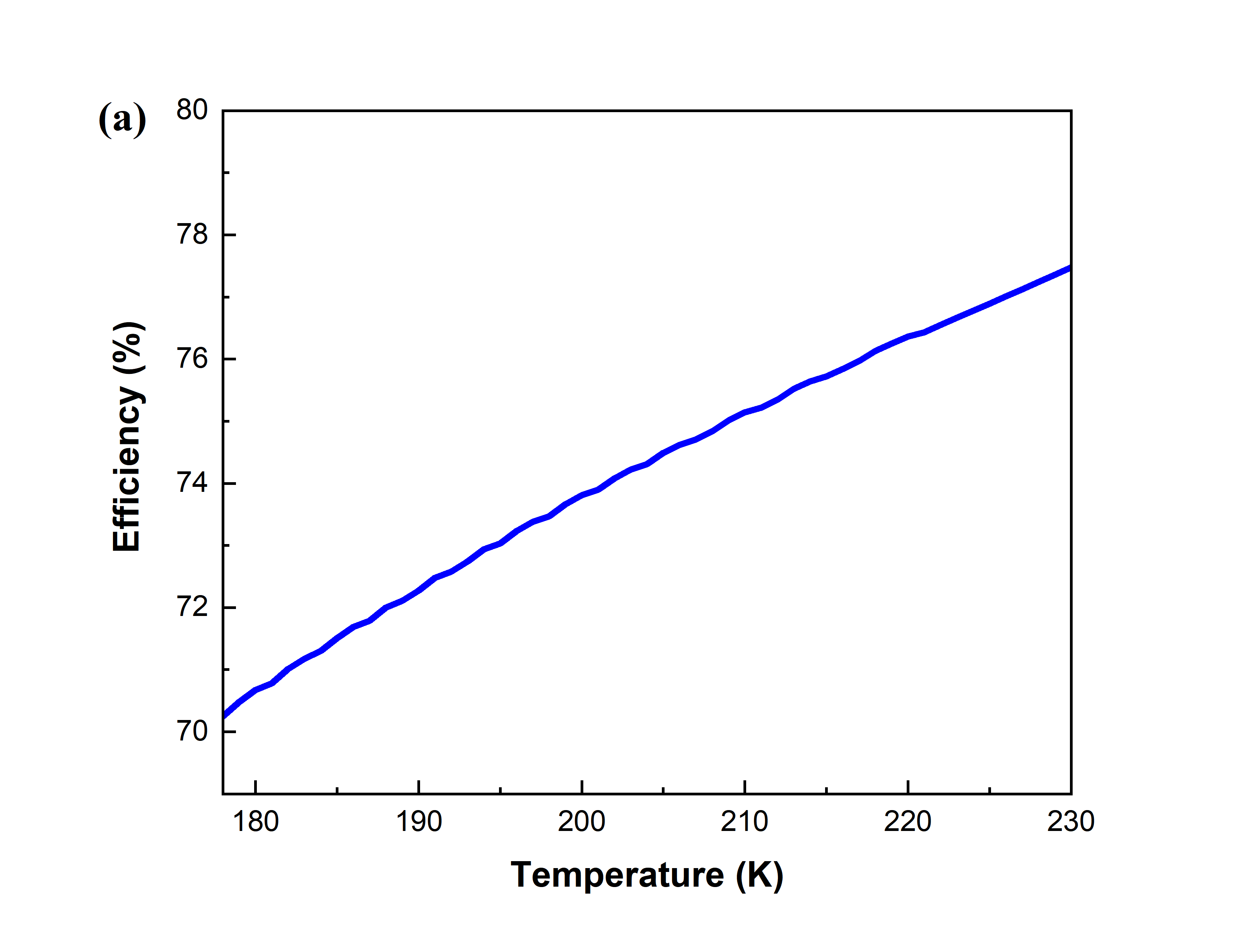}
\end{figure}

\begin{figure}[H]
\centering
\includegraphics[width=0.6\textwidth]{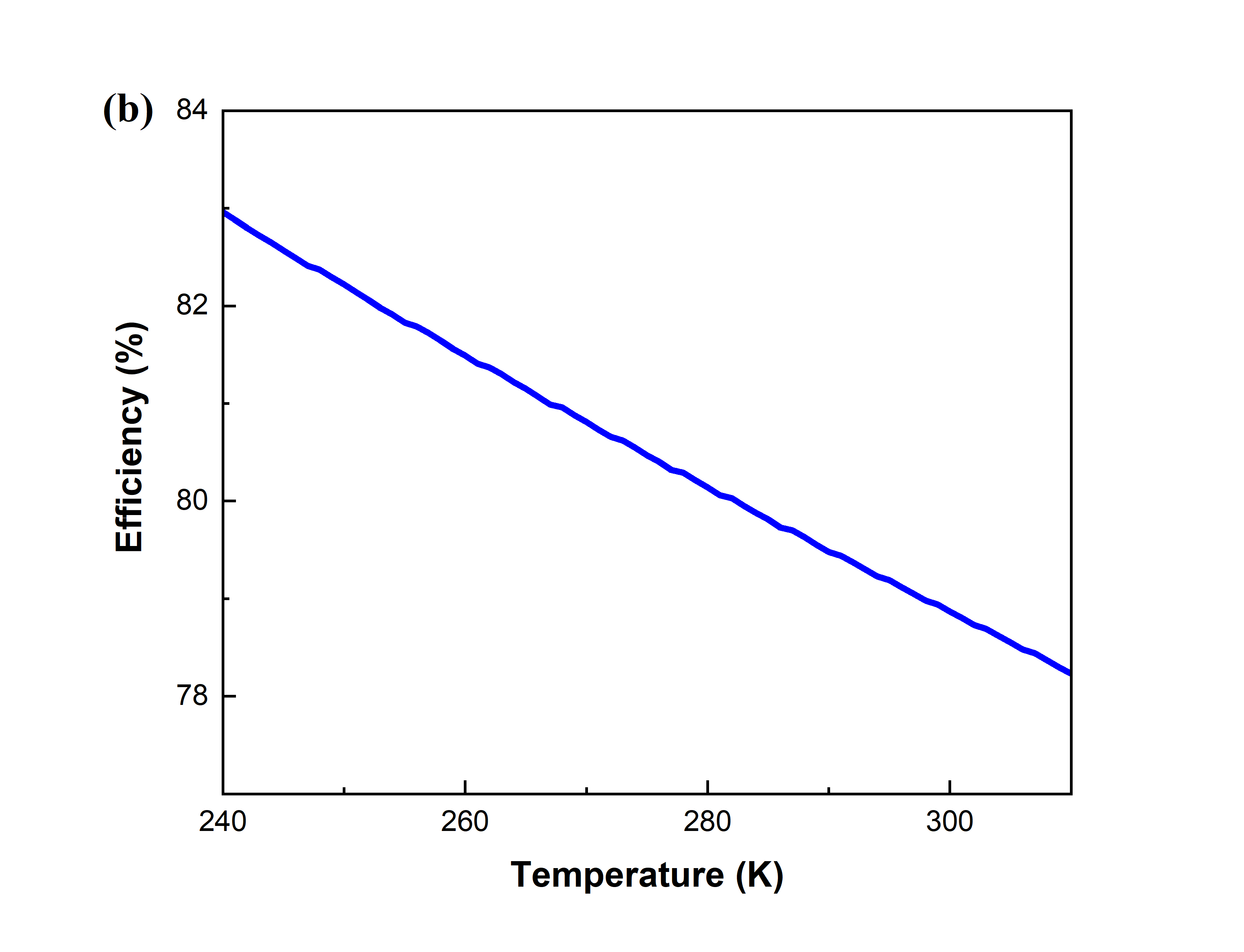}
\vspace{1em}
\caption{(a) Influence of the R14 temperature of the simulated cryostat outlet to the refrigeration efficiency of the system. (b) Influence of the R14 temperature of the compressor inlet to the refrigeration efficiency of the system.}
\label{fig16}
\end{figure}

\section{Conclusion}
\label{sec5}
An $R\&D$ experiment of the throttling refrigeration system for high cooling power recovery is designed and operated that utilizes R14 refrigerant to address the cooling power supply challenge for PandaX-xT online cryogenic distillation aimed at Rn removal. The system featured an innovative thermal synergistic circulation mechanism that overcomes the limitations of the traditional cooling methods, such as liquid nitrogen and GM cryocoolers. This results in energy savings and enables stable online distillation by increasing the distillation flow rate to 856 kg/h (5 LPM for LXe). The $R\&D$ experiment demonstrates stable system operation at an R14 flow rate of 0.2~kg/s, achieving a refrigeration capacity of 25~kW with an efficiency achieving 99\% $\pm$ 0.5\%. This performance translates to an annual savings of 3650~$m^3$ of liquid nitrogen and a reduction of approximately 211.5 kW in electrical consumption compared to using cryocoolers. Additionally, Aspen Hysys is employed for process simulation and optimization, achieving a simulation-experiment error of less than $3.64\%$. The study examines key parameters governing refrigeration capacity and efficiency, including the pressure drop within the throttle valve, pre-throttling temperature, pre-throttling pressure, the R14 temperature at the simulated cryostat outlet and the R14 temperature at the compressor inlet. These simulation outcomes inform the reconstruction and enhancement of the system to bolster the refrigeration efficiency. Furthermore, this study provides valuable experimental and simulation data that could aid in the design of similar throttling refrigeration systems for future applications, specifically addressing the core technical challenges of the PandaX-xT online distillation process for removing the Rn impurity.

\acknowledgments
The authors would like to thank the supports of the PandaX collaboration. 
This project is supported by grants from National Key R\&D Program of China (2023YFA1606204, 2023YFA1606200), National Science Foundation of China (Nos.52206015 and 12205189), grants from the Sichuan Science and Technology Program (No. 2025YFHZ0019), the Office of Science and Technology, Shanghai Municipal Government (Grant Nos. 23JC1410200, and 22JCJC1410200). We thank support from Double First Class Plan of the Shanghai Jiao Tong University, and the Tsung-Dao Lee Institute Experimental Platform Development Fund. We also thank the the sponsorship from the Chinese Academy of Sciences Center for Excellence in Particle Physics (CCEPP), Thomas and Linda Lau Family Foundation, Tencent and New Cornerstone Science Foundation in China, and Yangyang Development Fund. Finally, we thank the CJPL administration and the Yalong River Hydropower Development Company Ltd. for indispensable logistical support and other help.




\end{document}